\documentclass[twocolumn]{aastex631}
\pdfoutput=1

\newcommand\pdrtpy{{\tt pdrtpy}}
\newcommand\sphinx{{\tt sphinx}}
\newcommand\intensityunits {${\rm erg~cm^{-2}~s^{-1}~sr^{-1}}$}
\newcommand\fuvunits {${\rm erg~cm^{-2}~s^{-1}}$}
\received{September 2, 2022}
\revised{October 14, 2022}
\accepted{October 14, 2022}

\submitjournal{The Astronomical Journal}

\shorttitle{PhotoDissociation Region Toolbox}
\shortauthors{Pound and Wolfire}
\usepackage{listings}
\usepackage{xcolor}

\definecolor{codegreen}{rgb}{0,0.6,0}
\definecolor{codegray}{rgb}{0.5,0.5,0.5}
\definecolor{codepurple}{rgb}{0.58,0,0.82}
\definecolor{backcolour}{rgb}{0.95,0.95,0.92}

\lstdefinestyle{mystyle}{
    basicstyle=\ttfamily\footnotesize,
    breakatwhitespace=false,         
    breaklines=true,                 
    captionpos=b, 
    keepspaces=true,                 
    numbers=left,                    
    numbersep=5pt,                  
    showspaces=false,                
    showstringspaces=false,
    showtabs=false,                  
    tabsize=2,
    frame = shadow,
}

\begin{document}

\author[0000-0002-7269-342X]{Marc W. Pound}
\affiliation{Astronomy Department\\
University of Maryland\\
College Park, MD 20742}
\email{mpound@umd.edu}

\author[0000-0003-0030-9510]{Mark G. Wolfire}
\affiliation{Astronomy Department\\
University of Maryland\\
College Park, MD 20742}
\email{mwolfire@umd.edu}

\title{The PhotoDissociation Region Toolbox: Software and Models for Astrophysical Analysis}

\begin{abstract}
The PhotoDissociation Region Toolbox provides comprehensive, easy-to-use, public software tools and models that enable an understanding of the interaction of the light of young, luminous, massive stars with the gas and dust in the Milky Way and in other galaxies. It consists of an open-source Python toolkit and photodissociation region models for analysis of infrared and millimeter/submillimeter line and continuum observations obtained by ground-based and sub-orbital telescopes, and astrophysics space missions. 

Photodissociation regions (PDRs) include all of the neutral gas in the ISM where far-ultraviolet  photons dominate the chemistry and/or heating. In regions of massive star formation, PDRs are created at the boundaries between the H II regions and neutral molecular cloud, as photons with energies 6 eV $ < h \nu < $ 13.6 eV photodissociate molecules and photoionize metals. The gas is heated by photo-electrons from small grains and large molecules and cools mostly through far-infrared fine-structure lines like [O I] and [C II].

 The models are created from state-of-the art PDR codes that includes molecular freeze-out; recent collision, chemical, and photo rates; new chemical pathways, such as for oxygen chemistry; and allow for both clumpy and uniform media. The models predict the emergent intensities of many spectral lines and FIR continuum. The tools find the best-fit models to the observations and provide insights into the physical conditions and chemical makeup of the gas and dust. The PDR Toolbox enables novel analysis of data from telescopes such as ISO, Spitzer, Herschel, STO, SOFIA, SWAS, APEX, ALMA, and JWST.

\end{abstract}

\keywords{Photodissociation regions, Astronomy software, Molecular gas, Interstellar atomic gas}

\section{Introduction} \label{sec:intro}

Over twenty years ago, we created the Photodissociation Region Toolbox (PDRT), a web-based interface which allowed users to analyze the line and continuum emission from photodissociation regions (PDRs) \citep{Pound2008}. Back then, web programming meant Common Gateway Interface (CGI) and Perl was the workhorse scripting language.   Single pixel detectors were cutting edge technology and the sub-mm window had just begun to be explored.  We put together PDRT with Perl, HTML, Apache 1.3, FITS files, Concurrent Versioning System, shell scripts, and a sense of humor.  PDRT became a leading on-line site for analyzing PDRs and developed an international user base with users in over 35 countries. It garnered many refereed citations and the output plots have been used directly in published papers, and in posters and presentations, 
As new telescopes arrived, we added spectral lines and low metallicity models.   Browser-free scripting interfaces were created by users.  Although funding for the project ran out, we continued to add spectral lines when users requested and kept the service running.  Single pixel detectors gave way to cameras and the sub-mm science matured.  Now, thanks to renewed funding, we have rebuilt PDRT as an open-source Python 3 package called \pdrtpy\ with far more capability than the original CGI scripts. 
The PDR Toolbox website\footnote{\url{https://dustem.astro.umd.edu}} remains the central clearinghouse to keep users apprised of our work, with the code now moved to github.  The \pdrtpy\ version at the time of this writing is 2.2.9.

In this paper, we describe the scientific motivation to develop \pdrtpy, its architecture, and primary capabilities.  In Section \ref{sec:pdrs}, we discuss the astrophysics of PDRs. Section \ref{sec:philosophy} describes our development paradigm. 
In Section \ref{sec:models}, we describe the modeling physics and code. In Section \ref{sec:capabilities} we review \pdrtpy's core capabilities and in Section \ref{sec:future} we outline the development we would like to pursue in the near future.  To improve readability, example code listings are given in the Appendix rather than in the main text. The code listings are downloadable\footnote{Digital Repository at the University of Maryland identifier \url{http://hdl.handle.net/1903/29105 }} and can be used to reproduce Figures \ref{fig:modelplot}-\ref{fig:h2opr} in this paper.

\section{The Importance of PDRs}\label{sec:pdrs}
The interstellar medium (ISM) plays a central role in the lifecycle of
stars and galaxies.  The coldest phases
of the ISM, the molecular clouds, give rise to
star formation.  Stars return energy to their surroundings in
the form of photons and kinetic energy from winds and supernovae
explosions. In addition, stars enrich the ISM with metals that
affect the gas cooling.  Thus, understanding 
the production, chemistry, thermal balance and evolution
of the ISM  is
essential to understanding star formation and the evolution of
galaxies.
    
The infrared line and continuum emission from atoms, molecules,
and dust provide the observational diagnostics of the  stellar feedback.
The line and continuum emission arise primarily from PDRs. 
Classical PDRs are largely-neutral regions that are photodissociated and
heated by far-ultraviolet (FUV; 6 eV $<  h\nu  < 13.6$ eV) photons from nearby
massive stars \citep{TielensHollenbach1985, Wolfire2022}.
The same physical and chemical
processes that dissociate, partially ionize, and heat the gas in these 
classical PDRs are also important for the ISM as a whole, and PDRs are
now generally taken to include all regions where FUV photons play an
important role in the chemistry and/or heating of the gas. These regions
include diffuse and translucent H~I clouds and the warm neutral medium
in the ISM \citep{Wolfire2003},
the surfaces of molecular clouds
illuminated by the ambient radiation field and by nearby stars 
\citep{Kaufman2006}, the warm dust envelopes surrounding newly formed stars
\citep[e.g.,][]{Visser2012}, and  the neutral ISM in 
the disk and nuclei of normal and starburst galaxies 
\citep{Roussel2007,Kennicutt2011,Herrera-Camus2015}. Most of the non-stellar baryons within galaxies are in PDRs.

While gaining a conceptual understanding of PDRs is relatively easy, extracting the physical conditions and underlying physics from the observations is difficult.  For example, observers often use a simple large velocity gradient (LVG) model 
to analyze line ratios. Such a model is appropriate for emission from large clouds where there is an overall velocity gradient but not appropriate for individual star forming regions where PDR lines arise. In addition, LVG models typically do not account for the temperature distribution in thermal equilibrium  and abundance profiles in chemical balance through the emitting layer.
See \cite{Wolfire2022} for a review citing additional examples of PDRs, PDR models, and their applications. 

\section{Development Philosophy and Framework}\label{sec:philosophy}
Our approach is open-source development; user-friendliness; sensible and consistent interfaces; good documentation and examples; and responsive user support.   We use a Github open-source public repository\footnote{\url{https://github.com/mpound/pdrtpy}} that includes the Python code, text documentation files, and the FITS files of the models. The model FITS files can also be browsed and downloaded from the PDR Toolbox website.  Using Github Actions at code check-in, code is checked against the PEP8 coding style\footnote{\url{https://peps.python.org/pep-0008/}}, regression and integration tests are run, and code coverage of tests is calculated.  A separate repository contains Jupyter\footnote{\url{https://jupyter.org}} notebooks\footnote{\url{https://github.com/mpound/pdrtpy-nb}} that demonstrate how to do analysis with \pdrtpy.  (Notebooks are convenient but not required).  The repositories are governed by a GPL3 license.  We make use of major Python libraries such as {\tt astropy} \citep{astropy:2013, astropy:2018}, {\tt lmfit-py},\citep{lmfit}, {\tt matplotlib}, \citep{matplotlib2007CSE}, {\tt numpy} \citep{numpy2011}, {\tt scipy} \citep{jones2001scipy}, and {\tt emcee} \citep{Foreman_Mackey_2013}.  The full list of dependencies is given in the repository. \pdrtpy\  is installed via {\tt pip} or by cloning the git repository. We use \sphinx\footnote{\url{https://www.sphinx-doc.org/}} to generate documentation from code comments and text files which is then hosted on Read The Docs\footnote{\url{https://pdrtpy.readthedocs.io}}. Where applicable, we make an effort to ``promote'' keywords to our APIs, from e.g., {\tt matplotlib} and {\tt astropy}, that many users will already be familiar with.  This can be especially helpful in creating plots where users may want more fine-grained control than our default plots, which we already strive to make publication quality.

The fitting tools in the Toolbox inherit from a common parent class {\tt ToolBase}, which defines a few common attributes and properties and the {\tt run()} interface which all child classes must implement themselves.  The workflow for the user is to instantiate a fitting tool, optionally set some attributes, and invoke {\tt run()} to perform the fit.   Subsequently, they instantiate the companion plotting tool, which similarly derives from a parent {\tt PlotBase}, to explore the fit results. 

As astronomers who research PDRs, we have a reasonable idea of the kinds of tools that users need but welcome suggestions for desired functionality.   For example, model phase-space plotting with data overlay was requested by members of the SOFIA FEEDBACK \citep{Schneider2020} team so we prioritized its development and worked with them to beta test and to refine its functionality. It found immediate use in publications and talks \citep{Tiwari2022b,Tiwari2022c}.   

\section{The PDR models and model codes}\label{sec:models}

In the \pdrtpy\ distribution, all models are precomputed from PDR model codes, currently either the ``Wolfire-Kaufman" code which we have developed or the KOSMA-tau code \citep{Rollig2013,Rollig2022}. The models are computed using a given set of parameters (Table \ref{tab:modelparams}) and presented as grids of intensity or intensity ratio as a function of hydrogen nucleus density $n$ and radiation field strength $F_{FUV}$.  The results, collectively called a {\tt ModelSet}, are stored as FITS images in subdirectories organized by modelling code origin and major parameters such as metallicity.  A list of the available {\tt ModelSets} are given in Table \ref{tab:currentmodels}.

\begin{figure}
     \centering
     \includegraphics[width=\columnwidth,trim=2cm 6cm 2cm 2cm,clip]{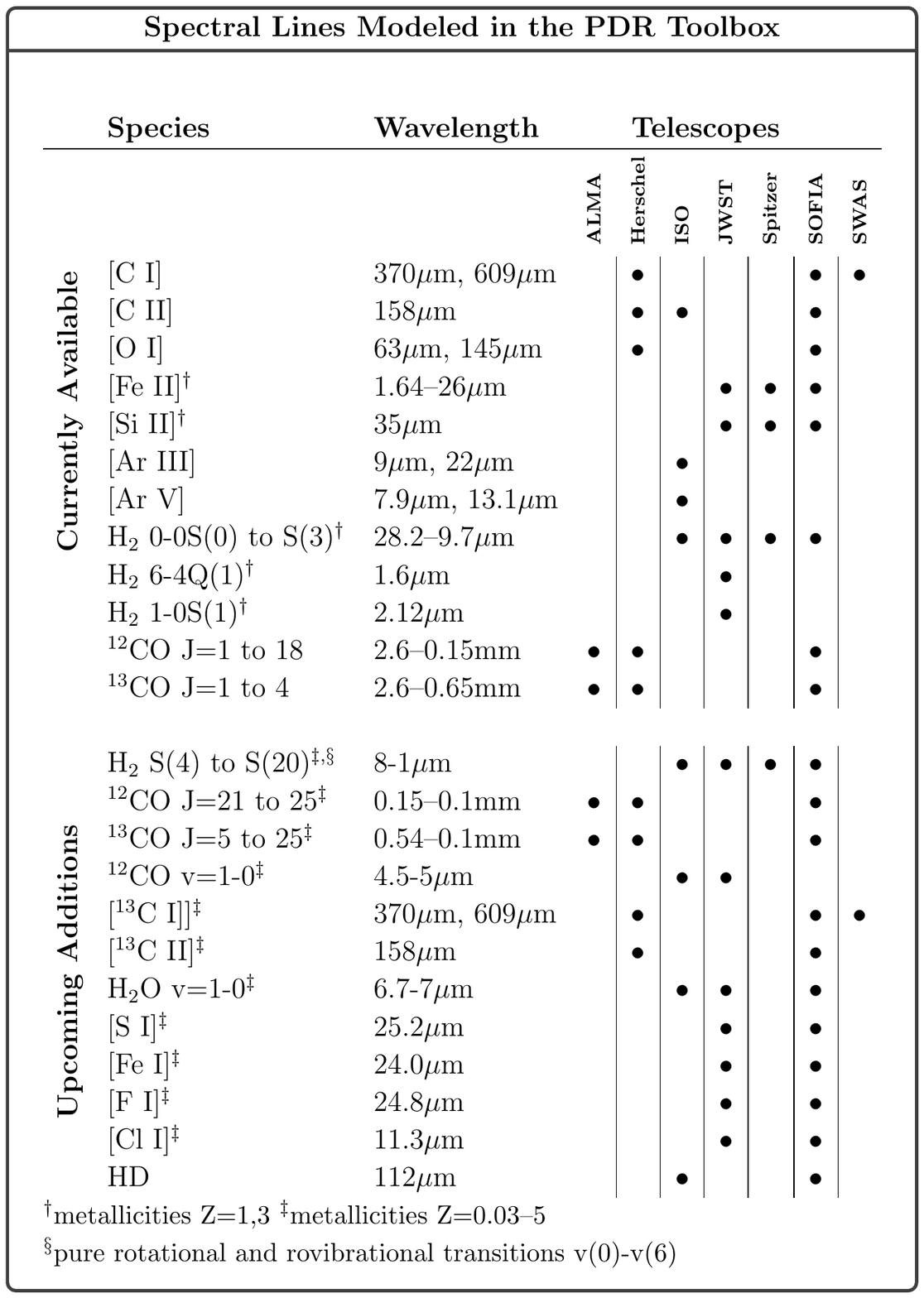}
     \caption{The spectral lines and metallicities currently available in PDRT and the upcoming additions. The Species column lists the spectral line designation; Wavelength gives the rest wavelength range covered by the models for the line(s). A dot in a Telescope column means that spectral line is observable with a given telescope (not including lines highly redshifted into the telescope bands).}
     \label{fig:pdrtlines}
 \end{figure}

The current set of distributed models, both Wolfire-Kaufman and KOSMA-tau, are most appropriate for the ``classical" PDRs
described in \cite{Wolfire2022} where stars illuminate nearby molecular clouds. The maximum depths from the cloud surface are
larger than found in diffuse or translucent molecular clouds
where $A_V{\sim} 1-2$ and the illumination is only
on the front side where for diffuse clouds the illumination would be on both the front
and back sides. Although a soft ($E < 100$ eV) X-ray spectrum is included in the Wolfire-Kaufman PDR code, neither set of PDR models
are appropriate for gas illuminated by hard X-ray radiation as would be emitted by an AGN (see also \citealt{Wolfire2022} for a comparison of PDR and X-ray dominated models).  The models cover a wide range of spectral lines that can be observed by many different telescopes (Figure \ref{fig:pdrtlines}). The telescopes listed in Figure \ref{fig:pdrtlines} are not an exhaustive list. Other telescopes such as APEX, KOSMA, AST/RO, and HHT have observed CO and CI, and high-$z$ spectral lines can be redshifted into observable bands of existing instruments.

 The Wolfire-Kaufman PDR model code based on the work of \cite{TielensHollenbach1985}
but with many updates since the early versions. It assumes a plane-parallel
geometry with a UV radiation field, cosmic-rays, and soft X-rays incident
on one side.
The main input parameters are the
radiation field strength in units of Habing \citep{Habing1968} fields ($G_0$) and a constant hydrogen
nucleus density $n$. Alternatively, 
the density can be derived self-consistently
from an input pressure. The code finds the gas temperature in thermal
equilibrium and abundances of atomic and molecular species in chemical balance.
The non-LTE level populations are calculated for the dominant coolants and
the emitted line intensities are found using an escape probability formalism.
Updates to the code are described  in \cite{Wolfire2010, Kaufman2006, Hollenbach2012} and  \cite{Neufeld2016}. 
More recent updates, and in particular those included in the ``wk2020''	models
feature	the photorates and dependence with depth
as given in \cite{Heays2017}, ${\rm ^{13}C}$ chemistry and line
emission, and O	collision rates	from \cite{Lique2018} as given in the MOLCAT \citep{Schoier2005} database.

In addition to hydrogen density and radiation field strength a large number of 
parameters could potentially be varied including the
gas phase metallicity, the abundance of grains and large
molecules, the microturbulent line-width, and the PDR depth.
The values for these parameters are given in Table~\ref{tab:modelparams} and are discussed in more
detail in the previous papers. Note that in the ``wk2020"
models we adopt a lower PDR depth ($A_{V,max} =7$) compared
to previous models to avoid possible time dependent effects in
the deeper layers. We also turn off chemistry on grain surfaces -- a constraint that will be lifted in future models.   Similarly, sets of models with low $A_V$ appropriate to diffuse or translucent clouds can be computed from this PDR model code.

The KOSMA-tau models and PDR model code are described in \cite{Rollig2013} and \cite{Rollig2022}. The geometry of the KOSMA-tau models differ from those of Wolfire-Kaufman. Instead of a plane-parallel geometry, KOSMA-tau uses an ensemble of spherical clumps with a spectrum of masses (``clumpy'') or a single clump (``non-clumpy''). Further, whereas 
the Wolfire-Kaufman code has a fixed incident spectral
energy distribution (that of the interstellar radiation field) and grain model (interstellar medium grains with $R_V=3.1$), the KOSMA-tau code  can independently 
vary them (see Table \ref{tab:currentmodels}).
We were provided with FITS files of model spectral line intensity ratios and intensities by the KOSMA-tau authors for use in PDRT. 

The choice of PDR model can have significant effects on the predicted line intensities (see Figure \ref{fig:modelplot}). This can give physical insight into the PDR conditions, for instance, whether the data are better represented by a clumpy or plane-parallel medium. 

\begin{deluxetable*}{lllll}
 \tablecaption{Example Parameters of PDR Models\label{tab:modelparams}}
 \tablewidth{0pt}
 \tablehead{
 \colhead{Parameter} & \colhead{units} & \colhead{Symbol} & \colhead{WK2020} & \colhead{KOSMA-tau} 
 }
 \decimalcolnumbers
 \startdata
    Carbon abundance          &  &$X_C$       & $1.6\times 10^{-4}$ & $2.34\times 10^{-4}$\\ 
    $^{13}$C abundance        &  &$X_{13C}$   & $3.2\times 10^{-6}$ & $3.2\times 10^{-6}$\\
    Oxygen abundance          &  &$X_O$       & $3.2\times 10^{-4}$ & $4.47\times 10^{-4}$\\ 
    $^{18}$O abundance        &  &$X_{18O}$   & \nodata & $8.93 \times 10^{-7}$ \\
    Silicon abundance         &  &$X_{Si}$    & $1.7\times 10^{-6}$ & $3.17\times 10^{-6}$\\ 
    Sulfur abundance          &  &$X_{S}$     & $2.8\times 10^{-5}$ & $7.41\times 10^{-6}$\\ 
    Iron abundance            &  &$X_{Fe}$    & $1.7\times 10^{-7}$ & $1.0\times 10^{-6}$\\ 
    Magnesium abundance       &  &$X_{Mg}$    & $1.1\times 10^{-6}$ & $3.2\times 10^{-6}$ \\ 
    Nitrogen abundance        &  &$X_N$       & 0 & $8.32\times 10^{-5}$ \\
    Fluorine abundance        &  &$X_F$       & $1.8 \times 10^{-8}$ & $6.68 \times 10^{-9}$\\
    Helium abundance          &  &$X_{He}$   & $ 0.1 $              & $8.51 \times 10^{-2}$ \\
    PAH abundance             &  &$X_{PAH}$   & $2\times 10^{-7}$ & \nodata\tablenotemark{a}\\ 
    Dust and Metals with respect to local ISM & & Z & 1 & 1\\
    Dust abundance relative to diffuse ISM &  & $\delta_d$         & 1 & 1\\ 
    FUV dust opacity/visual extinction  &  &$\tau_{FUV}/\tau_V$ & 1.8 & \nodata\tablenotemark{b}\\
    Maximum optical depth             &       & $A_{V,max}$ & 7 & \nodata\tablenotemark{c} \\
    Dust visual extinction per H atom      & cm$^{-2}$  &$\sigma_V$ & $5.26\times 10^{-22}$  & \nodata\tablenotemark{d}\\
    Formation rate of H$_2$ on dust        &  s$^{-1}$ &$R_{form}$ & $6\times 10^{-17}$  & \nodata\tablenotemark{e} \\ 
    Turbulent Doppler velocity& km s$^{-1}$ & $\delta v_{D}$ & 1.5     & \nodata\tablenotemark{f} \\
    Cosmic ray ionization rate per H nuclueus & s$^{-1}$ & $\zeta_{CR}$ & $2.0\times 10^{-16}$
    \tablenotemark{g}& $2.0 \times 10^{-16}$ \\    
    Cloud H density                        & cm$^{-3}$ & $n$ & $10^{1} - 10^{7}$ & $10^{3} - 10^{7}$\tablenotemark{h} \\ 
    Incident UV flux\tablenotemark{i} & \fuvunits & $F_{FUV}$ & $10^{-3.3} - 10^{3.7}$  &  $10^{-2.5} - 10^{3.4}$\\ 
 \enddata
 \tablenotetext{a}{Following the \cite{Weingartner2001} prescription, the PAH abundance is not specified.}
 \tablenotetext{b}{Depends on dust model, see Table 4 of \cite{Rollig2013}.}
\tablenotetext{c}{Depends on mass and density of model.}
 \tablenotetext{d}{Depends on dust model: WD01-7: $5.24 \times 10^{-22}$, WD01-21: $5.05 \times 10^{-22}$, WD02-25: $4.88e\times 10^{-22}$.}
 \tablenotemark{e}{Computed following \cite{Cazaux2004,Cazaux2010}.}
 \tablenotetext{f}{Doppler velocity computed from \citep{Larson1981} mass-line width  relation.}
 \tablenotetext{g}{Assumes the ionization rate falls as $\zeta_{CR}/(1+N/1.0\times 10^{21}\,{\rm cm^{-2}})$ with a minimum of $2.0\times 10^{-17}$ ${\rm s^{-1}}$.}
 \tablenotetext{h}{This is the density at the surface. KOSMA-tau assumes a certain profile, typically leading to a central density $\sim 11$ times higher and a mean density that is $\sim 1.9$ times the surface density.}
 \tablenotetext{i}{\cite{Draine1978} spectral energy distribution.}
 \end{deluxetable*}  
 
    \section{Capabilities}\label{sec:capabilities}
    \subsection{Data Representation}\label{subsec:Data}
    Observations in \pdrtpy\ are represented by the {\tt Measurement} class. A {\tt Measurement} consists of a value and an error. These can be single-valued, an array of values, or an image and can be in intensity units (equivalent to \intensityunits) or in ${\rm K~km~s^{-1}}$ which is typical of mm/submm spectral line observations. In the typical case of an image observation, the {\tt Measurement} is a representation of a FITS file with two header data units (HDUs): the first HDU is the spatial map of intensity and the second HDU is the spatial map of the errors. An image-based {\tt Measurement} carries a world coordinate system and traditional FITS-like header.
     {\tt Measurement} is based on the {\tt CCDData} class of {\tt astropy} with additional properties such as beam parameters and support for arithmetic operators.  In arithmetic operations, errors and units are correctly propagated through the underlying {\tt astropy} code.  Users identify their {\tt Measurements} with one of \pdrtpy's predefined ID strings, e.g. ``CII\_158" for the [C II] 158 $\mu$m spectral line.  These identifiers are used by \pdrtpy\ to match observations with models which are similarly identified.  
    
\begin{figure}
    \centering
    \includegraphics[width=\columnwidth]{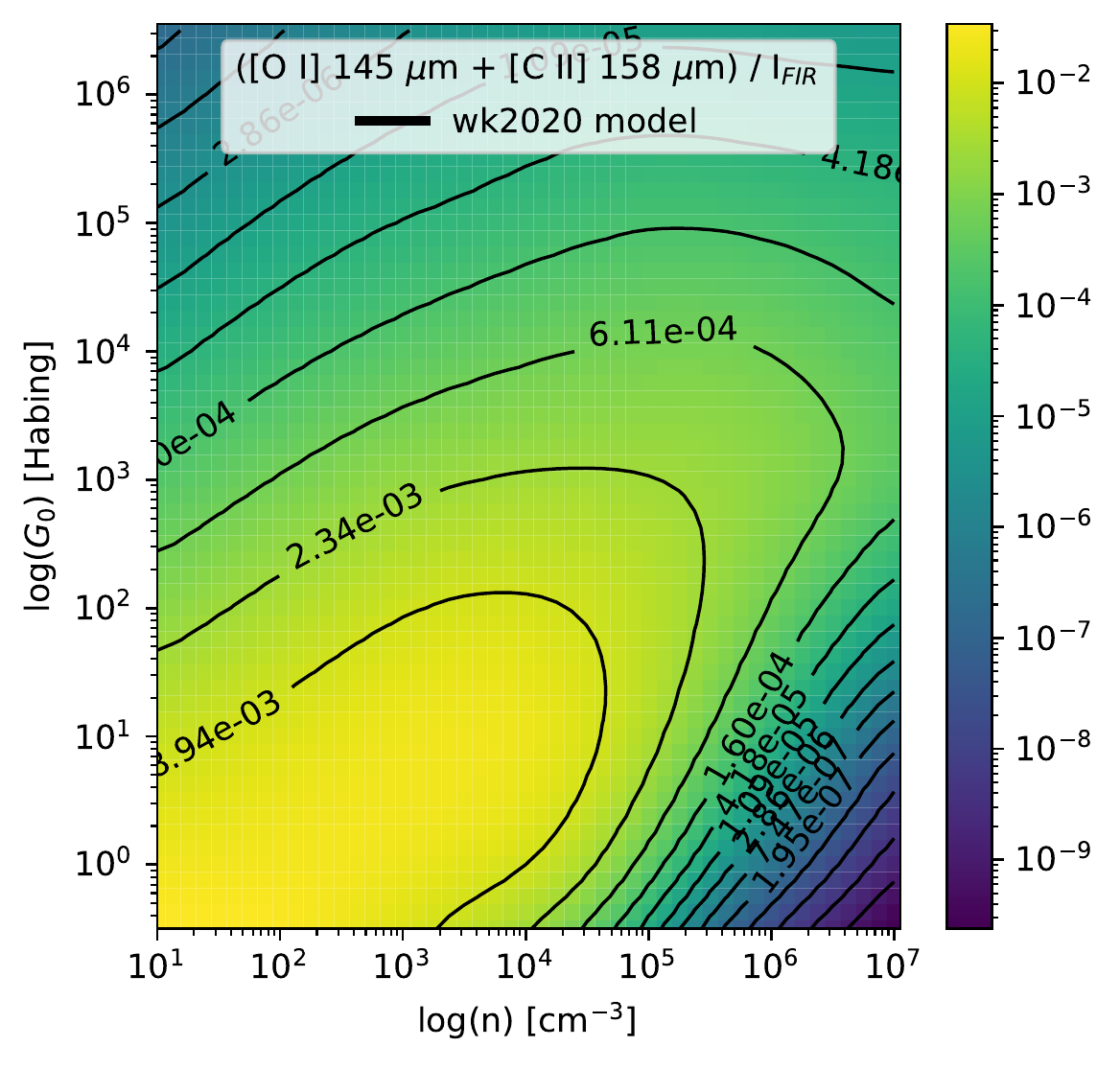}\\
    \includegraphics[width=\columnwidth]{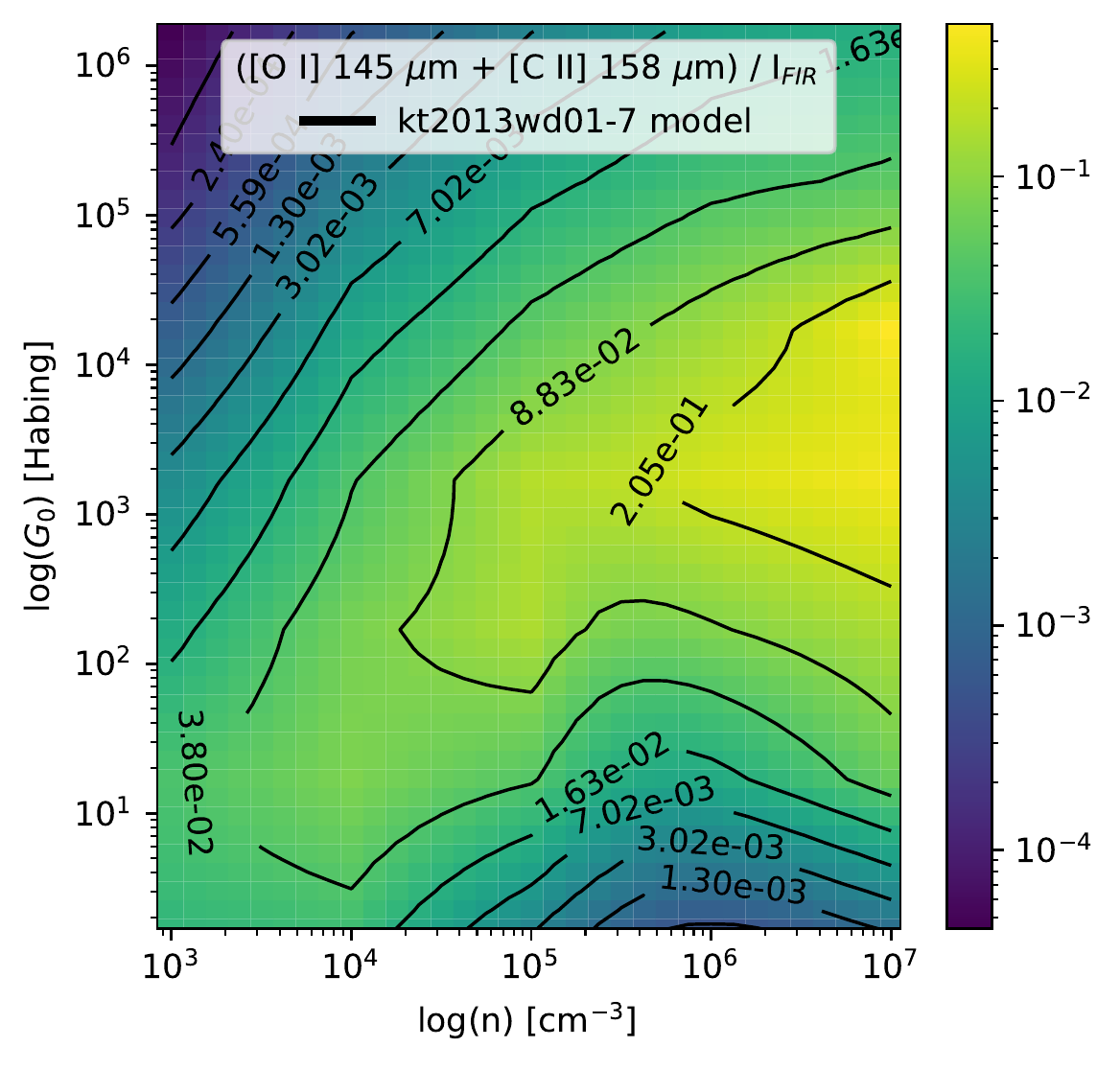}
    \caption{Examples of model plots. {\it top)} The WK2020 model for the ratio of the sum of the [O I] 145 $\mu$m and [C II] 158 $\mu$m intensities divided by far-infrared intensity integrated between 8 $\mu$m and 1 mm, $I_{FIR}$, computed as a function of H nucleus density $n$ and FUV field $G_0$.  {\it bottom)} The same model intensity ratio as computed in the kt2013wd01-7, clumpy, M=100 M$_{\odot}$, $R_V=3.1$ model. The user-friendly flexibility of \pdrtpy\ allows choice of Habing units for $G_0$, log normalization for the image intensities with a color-blind friendly colormap, and labelled black contours. See Listing \ref{listing1}. }
    \label{fig:modelplot}
\end{figure}

    The models in \pdrtpy\ are two-dimensional grids of either intensities or ratios of intensities as a function of hydrogen nucleus volume density (cm$^{-3}$) and incident FUV field, $ F_{FUV}$ (\fuvunits\ or equivalent). Since these are stored on disk as FITS images, they are also represented internally as {\tt Measurements}, but with no errors. Because of the built-in operator support, this makes straightforward arithmetic operations that involve both observations and models.   There are different conventions for the units of $F_{FUV}$ depending on different approximations to the local interstellar radiation field-- cgs units, Habing units \citep[1 Habing = $G_0$ = 1.63$\times 10^{-3}$ \fuvunits]{Habing1968}, Draine units \citep[1 Draine = $\chi$ = 2.72$\times 10^{-3}$ \fuvunits]{Draine1978}, and Mathis units \citep[1 Mathis = 1.81$\times 10^{-3}$ \fuvunits]{Mathis1983}; \pdrtpy\  defines each as an astropy {\tt Quantity} to convert seamlessly between them. This allows the user to, for instance, create plots in their preferred unit (See, e.g.,  Listing \ref{listing1} and Figure \ref{fig:modelplot}).
    
    Collections of models are managed by the {\tt ModelSet} class.  A {\tt ModelSet} represents a coherent collection of models that were created using the same modeling code and physical parameters (e.g., Table \ref{tab:modelparams}).  The list of currently available {\tt ModelSets} is given in Table \ref{t-modelsets}.  Users retrieve individual models from a {\tt ModelSet} using their identifiers.   Listing 
    \ref{listing1} shows an example of instantiating a {\tt ModelSet}, retrieving individual models from it, and plotting a model (Figure \ref{fig:modelplot}).

\begin{figure*}
    \centering
    \includegraphics[width=\textwidth]{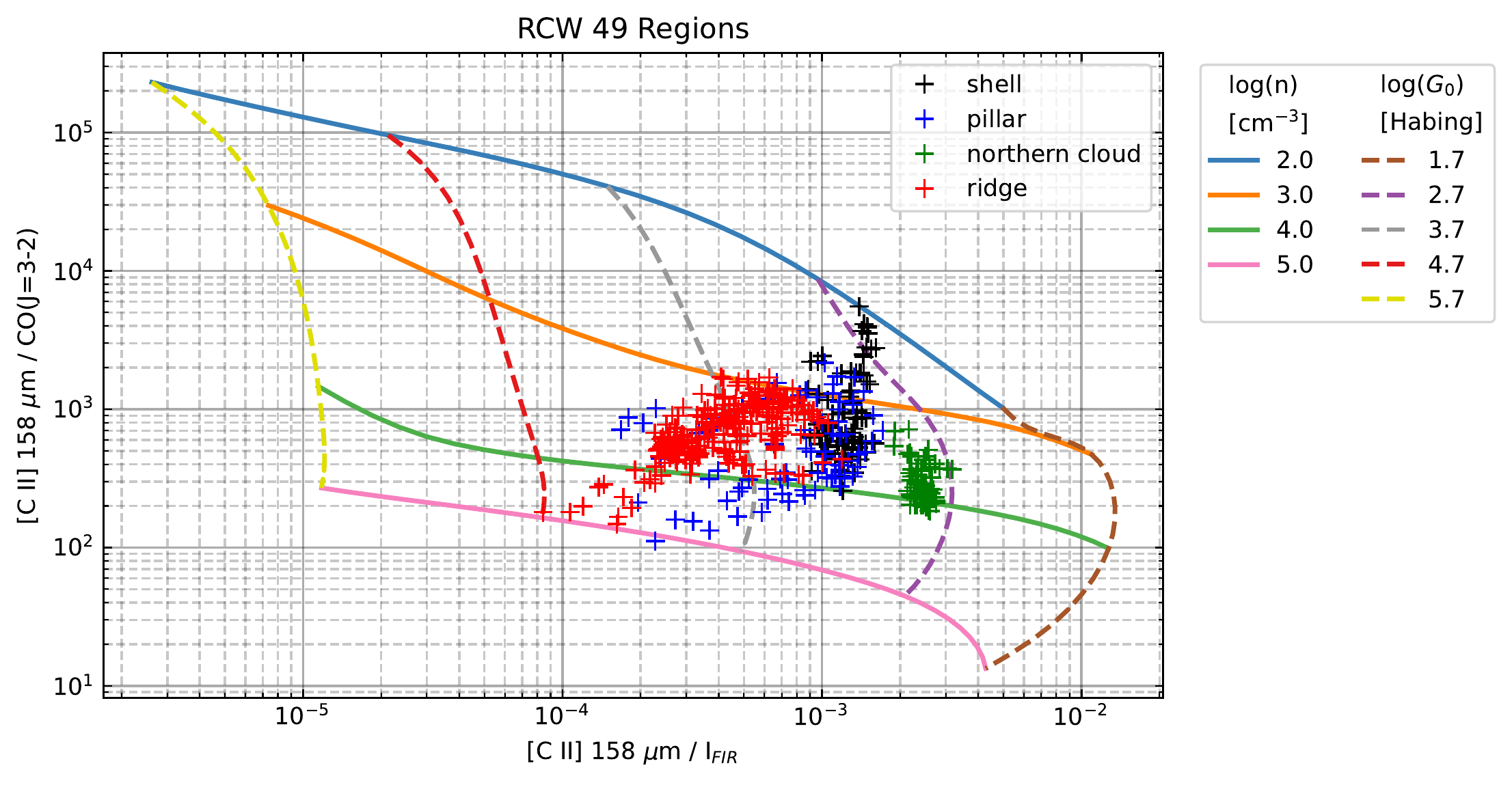}
    \caption{An example of a phase space plot showing lines of constant $n$ and $G_0$ as a function of spectral line intensity ratios. The plot uses WK2020 models and [C II], CO(3-2), and FIR data in the RCW 49 PDR from \cite{Tiwari2021}. The crosses are observed [C II]/FIR and [C II]/CO(3-2) intensity ratios for 4 regions in RCW 49. See Listing \ref{listing1}}
    \label{fig:phasespace}
\end{figure*}

    Another way to visualize the models is through a phase space diagram which can plot lines of constant $n$ and $F_{FUV}$ as a function of spectral line intensity or intensity ratio. Adding observed data to the plot lets the astronomer understand the conditions in different regions, as was done by \cite{Tiwari2021} for RCW 49 (Listing \ref{listing1}, Figure \ref{fig:phasespace}). Phase space diagrams can also be useful for making predictions of line strength or estimating density and radiation field when the user doesn't have enough observations to fit with {\tt LineRatioFit}.
   
   \begin{deluxetable*}{llrrrrl}\label{t-modelsets}
 \tablecaption{PDR Models Currently Supported\label{tab:currentmodels}}
 \tablewidth{0pt}
 \tablehead{
\colhead{PDR code} & 
\colhead{name}& 
\colhead{version} & 
\colhead{medium} & 
\colhead{Z} & 
\colhead{mass\tablenotemark{a}} & 
\colhead{$R_V$\tablenotemark{b}}  \\
\colhead{} & 
\colhead{} & 
\colhead{} & 
\colhead{} & 
\colhead{} & 
\colhead{$\mathrm{M_{\odot}}$} &
\colhead{}
}
\decimalcolnumbers
\startdata
Wolfire/Kaufman & wk2006 & 2006 & constant density & 1.0 & \nodata &  3.1 \\
Wolfire/Kaufman & wk2006 & 2006 & constant density & 3.0 & \nodata &  3.1 \\
Wolfire/Kaufman & wk2020 & 2020 & constant density & 1.0 & \nodata &  3.1 \\
Wolfire/Kaufman & smc\tablenotemark{c} & 2006 & constant density & 0.1 &\nodata  & 3.1 \\
Wolfire/Kaufman & lmc\tablenotemark{d} & 2020 & constant density & 0.5 & \nodata & 3.1 \\
KOSMA-tau & kt2013wd01-7  & WD01-7  & clumpy     & 1.0 & 100.0 & 3.1  \\
KOSMA-tau & kt2013wd01-7  & WD01-7  & non-clumpy & 1.0 & 100.0 & 3.1  \\
KOSMA-tau & kt2013wd01-21 & WD01-21 & clumpy     & 1.0 & 100.0 & 4.0  \\
KOSMA-tau & kt2013wd01-21 & WD01-21 & non-clumpy & 1.0 & 100.0 & 4.0  \\
KOSMA-tau & kt2013wd01-25 & WD01-25 & clumpy     & 1.0 & 100.0 & 5.5  \\
KOSMA-tau & kt2013wd01-25 & WD01-25 & non-clumpy & 1.0 & 100.0 & 5.5  \\
KOSMA-tau & kt2013wd01-7  & WD01-7  & clumpy     & 1.0 & 10.0  & 3.1  \\
KOSMA-tau & kt2013wd01-7  & WD01-7  & non-clumpy & 1.0 & 10.0  & 3.1  \\
KOSMA-tau & kt2013wd01-21 & WD01-21 & clumpy     & 1.0 & 10.0  & 4.0  \\
KOSMA-tau & kt2013wd01-21 & WD01-21 & non-clumpy & 1.0 & 10.0  & 4.0  \\
KOSMA-tau & kt2013wd01-25 & WD01-25 & clumpy     & 1.0 & 10.0  & 5.5  \\
KOSMA-tau & kt2013wd01-25 & WD01-25 & non-clumpy & 1.0 & 10.0  & 5.5  \\
KOSMA-tau & kt2013wd01-7  & WD01-7  & clumpy     & 1.0 & 1.0 & 3.1 \\
KOSMA-tau & kt2013wd01-7  & WD01-7  & non-clumpy & 1.0 & 1.0 & 3.1 \\
KOSMA-tau & kt2013wd01-21 & WD01-21 & clumpy     & 1.0 & 1.0 & 4.0 \\
KOSMA-tau & kt2013wd01-21 & WD01-21 & non-clumpy & 1.0 & 1.0 & 4.0 \\
KOSMA-tau & kt2013wd01-25 & WD01-25 & clumpy     & 1.0 & 1.0 & 5.5 \\
KOSMA-tau & kt2013wd01-25 & WD01-25 & non-clumpy & 1.0 & 1.0 & 5.5 \\
KOSMA-tau & kt2013wd01-7  & WD01-7  & clumpy     & 1.0 & 0.1 & 3.1 \\
KOSMA-tau & kt2013wd01-7  & WD01-7  & non-clumpy & 1.0 & 0.1 & 3.1 \\
KOSMA-tau & kt2013wd01-21 & WD01-21 & clumpy     & 1.0 & 0.1 & 4.0 \\
KOSMA-tau & kt2013wd01-21 & WD01-21 & non-clumpy & 1.0 & 0.1 & 4.0 \\
KOSMA-tau & kt2013wd01-25 & WD01-25 & clumpy     & 1.0 & 0.1 & 5.5 \\
KOSMA-tau & kt2013wd01-25 & WD01-25 & non-clumpy & 1.0 & 0.1 & 5.5 \\
KOSMA-tau & kt2013wd01-7  & WD015.5 & clumpy     & 1.0 & 1000.0 & 3.1 \\
KOSMA-tau & kt2013wd01-7  & WD01-7  & non-clumpy & 1.0 & 1000.0 & 3.1 \\
KOSMA-tau & kt2013wd01-21 & WD01-21 & clumpy     & 1.0 & 1000.0 & 4.0 \\
KOSMA-tau & kt2013wd01-21 & WD01-21 & non-clumpy & 1.0 & 1000.0 & 4.0 \\
KOSMA-tau & kt2013wd01-25 & WD01-25 & clumpy     & 1.0 & 1000.0 & 5.5 \\
KOSMA-tau & kt2013wd01-25 & WD01-25 & non-clumpy & 1.0 & 1000.0 & 5.5 \\
\enddata
\tablenotetext{a}{For clumpy models this is the maximum clump mass. For non-clumpy models, it is the mass of a single spherical clump.}
\tablenotetext{b}{KOSMA-tau models use dust properties from \cite{Weingartner2001d}. }
\tablenotetext{c}{Limited set of models for Small Magellanic Cloud.}
\tablenotetext{d}{Limited set of models for Large Magellanic Cloud.}
\end{deluxetable*}

    \subsection{Fitting Observations and Plotting Results}
    \subsubsection{Intensity Ratios}
    It has been shown that far-infrared line and continuum observations can be used to determine the physical properties of PDRs including the incident FUV radiation field, the gas density, and the surface temperature \cite{Wolfire1990,Kaufman1999,Kaufman2006}.  These authors showed that ratios of intensities are particularly effective for determining $n$ and $F_{FUV}$ since to first order the beam filling factors cancel\footnote{We note that for unresolved observation beam filling factors may not cancel and an additional correction to normalize the filling factors are needed. See the detailed procedure given in \cite{Kaufman2006} and additional caveats in comparing models with observations in \cite{Wolfire2022}.}.  
    
    In \pdrtpy, the {\tt LineRatioFit} tool takes intensity {\tt Measurements} and a {\tt ModelSet} as input, computes the intensity ratios that have entries in the {\tt ModelSet}, and finds the best-fit $n$ and $F_{FUV}$.   The fit result matches the input --  single-value, array, or spatial image.  The available fitting algorithms are non-linear least squares minimization (NNLS) or the Monte Carlo Markov Chain (MCMC) to determine the posterior probability density function (PDF) of the fitted parameters.  Both are managed through {\tt lmfit} which capably delegates via easy-to-use high-level interfaces to {\tt scipy.optimize} for NNLS or {\tt emcee} for MCMC.  

\begin{figure*}
    \centering
    \includegraphics[width=\textwidth]{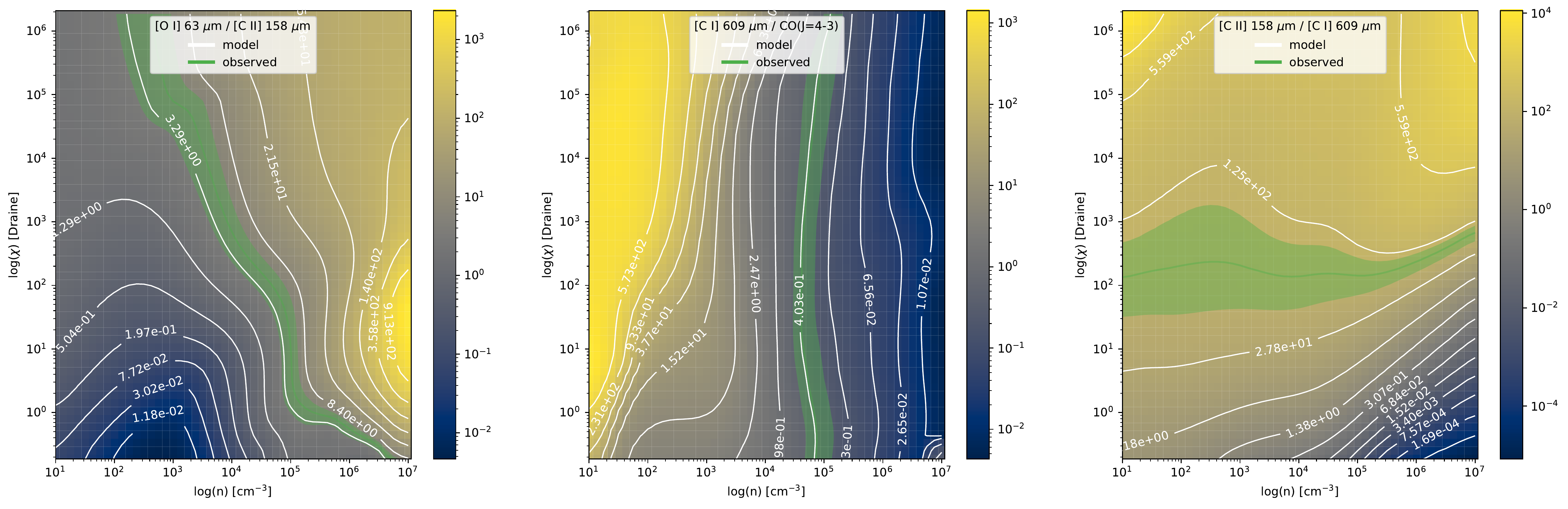}
    \caption{Plots created in Listing \ref{listing2} using the {\tt LineRatioPlot.ratios\_on\_models} method showing the observed ratios with errors overlayed on the matching models. The observational errors ($1 \sigma$) are shown as shaded regions around the solid observation line. Axis units, colors, contours, and other plot parameters can be modified by the user via the API. The data are values chosen for demonstration purpose.}
    \label{fig:ratios}
\end{figure*}

\begin{figure}
    \centering
    \includegraphics[width=0.45\textwidth]{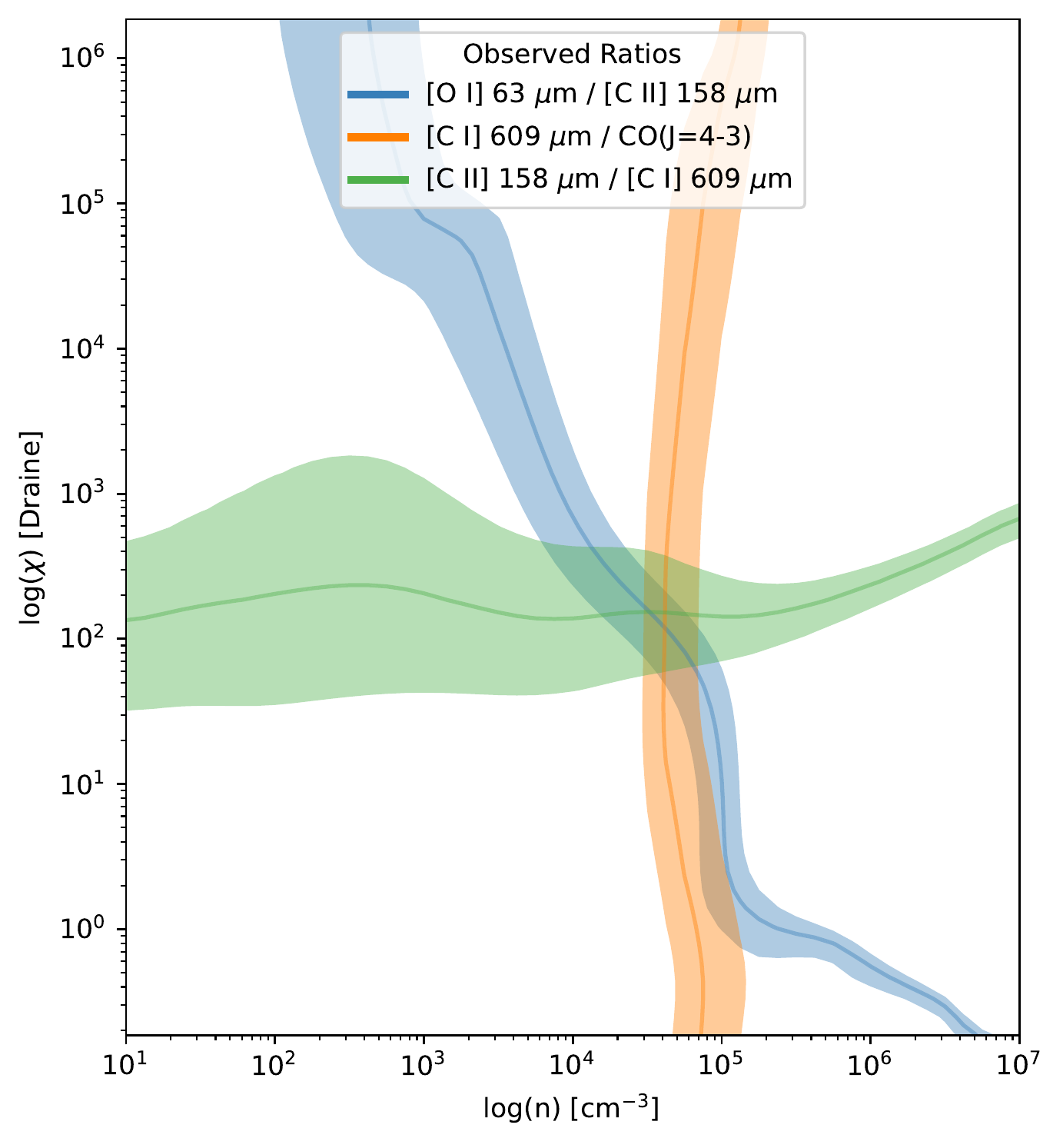}
    \includegraphics[width=0.45\textwidth]{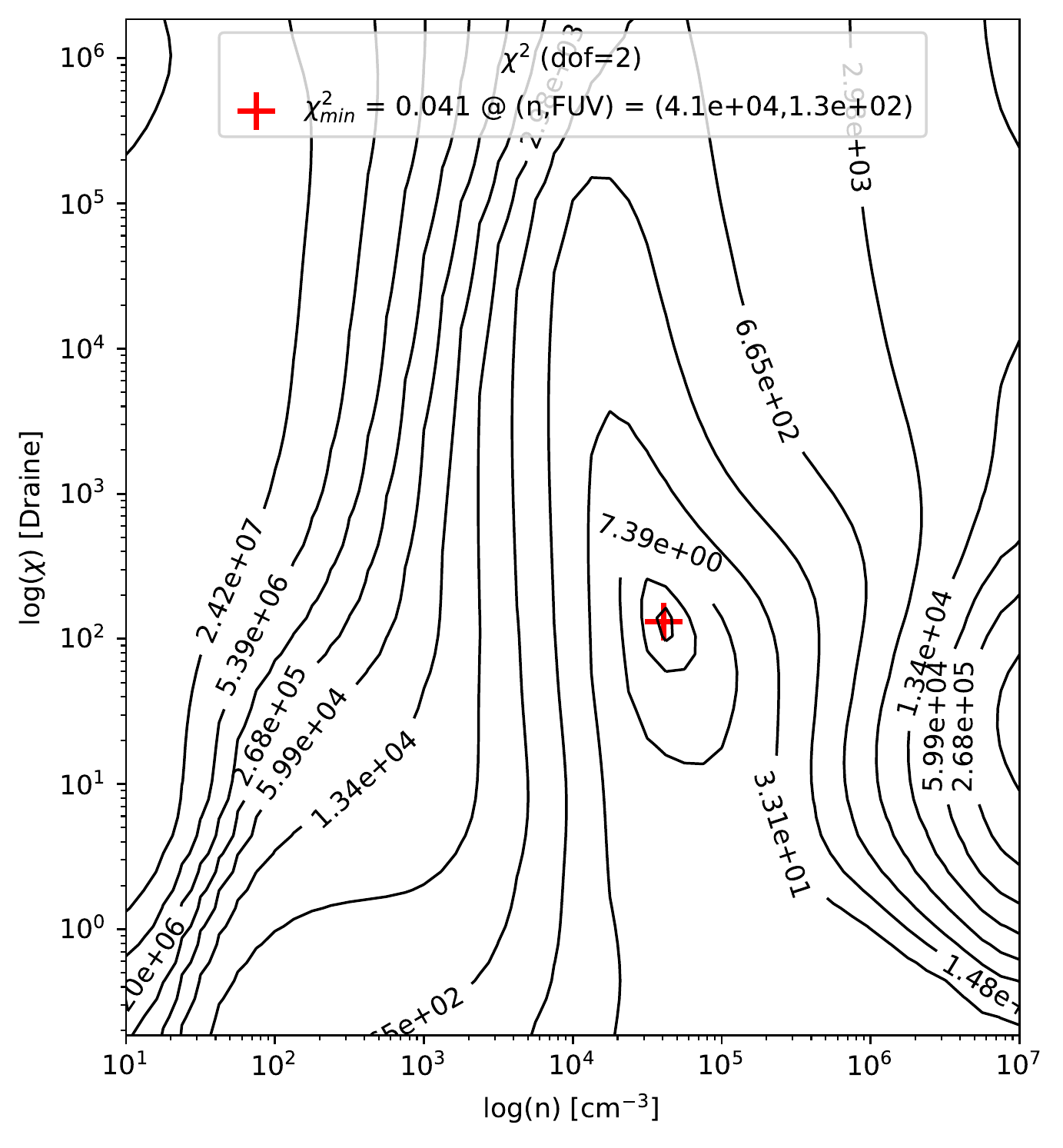}

        \caption{Plots created in Listing \ref{listing2} using the {\tt LineRatioPlot.overlay\_all\_ratios} and  {\tt LineRatioPlot.chisq} methods. {\emph top)} The observed ratios used in the fit overlayed on the model space; solid lines are observed values and shaded regions are $1 \sigma$ errors. The intersection of the lines indicates the region of model space where the most likely density and radiation field lie.  {\emph bottom)} Contour plot of reduced chi-square from the fit. Red cross indicates the minimum $\chi^2$ and best-fit density and radiation field. Plot parameters are modifiable by the user via the API.}
    \label{fig:chisq}
\end{figure}

\begin{figure}
    \centering
    \includegraphics[width=\columnwidth]{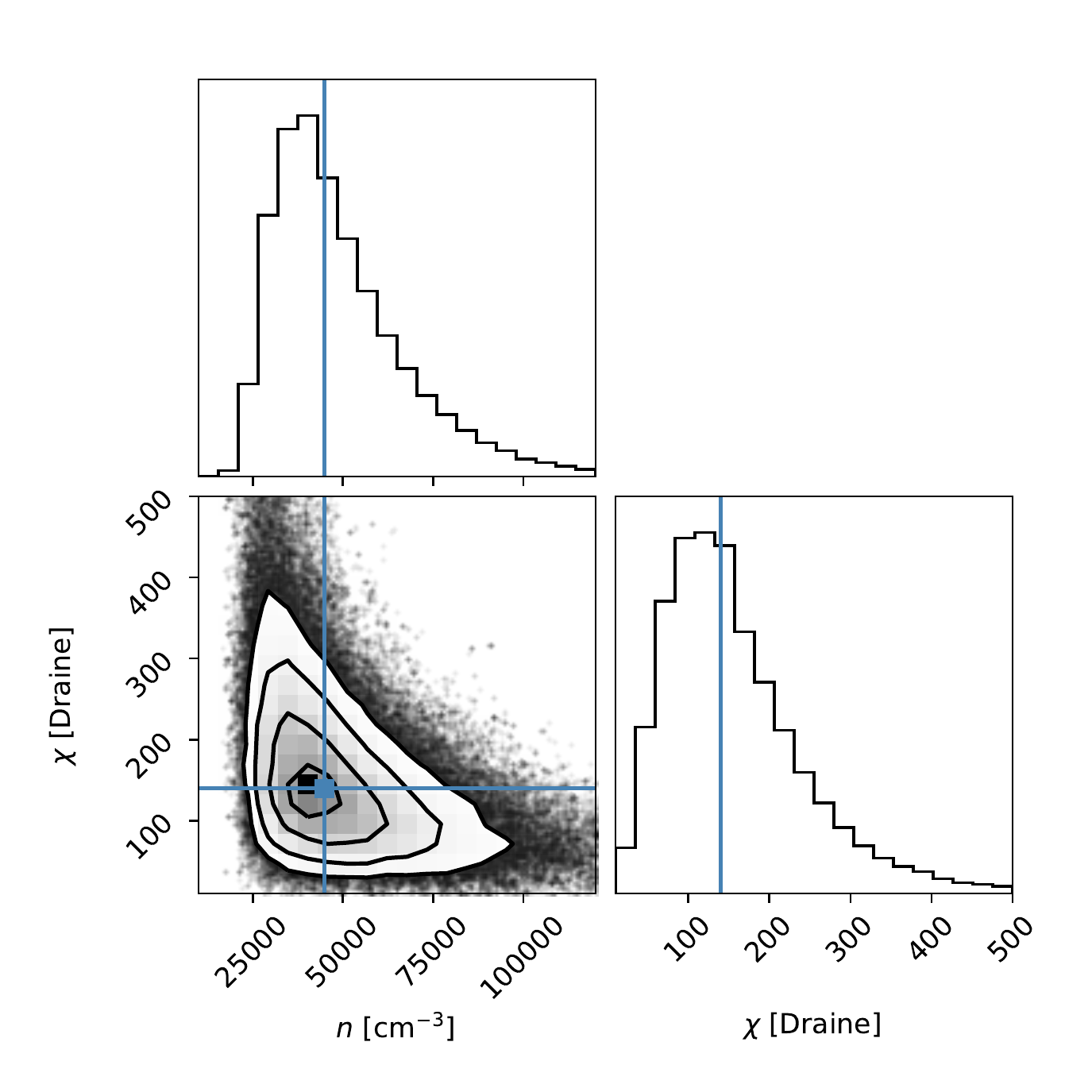}
    \caption{A corner plot showing the histograms probability density functions of density $n$ and radiation field $\chi$ from the MCMC fit in Listing \ref{listing3}. Blue lines indicate most probable values. }
    \label{fig:emcee}
\end{figure}

    Listing \ref{listing2} gives an example of determining  $n$ and $F_{FUV}$ from single-pixel (or single-beam) observations using {\tt LineRatiofit} and plotting the results with {\tt LineRatioPlot}.   Integrated intensity observations of [O I]~63 \micron,  [C I]~609 \micron,  CO(J=4-3), and [C II]~158 \micron\ are used to create three ratios and the {\tt run()} method invokes NNLS minimization to determine the best-fit quantities.  The results can be inspected with print statements, ratio plots (Figure \ref{fig:ratios}), overlay plots, and chi-square plots (Figure \ref{fig:chisq}). In Listing \ref{listing3}, we show how to fit $n$ and $F_{FUV}$ using MCMC and how to create the traditional corner plot with the desired axes (Figure \ref{fig:emcee}).

 Listing \ref{listing3} shows how to fit the {\tt Measurements} from Listing \ref{listing2}  using the MCMC method by passing the appropriate arguments to {\tt LineRatioFit.run()}.  The {\tt emcee} package is used and keywords specific to {\tt emcee} can be passed in, e.g., \textit{steps} indicates how many samples to draw from the posterior distribution for each walker.  Listing \ref{listing3} also shows how to create a custom corner plot of the results (Figure \ref{fig:emcee}) using the {\tt corner} package \citep{corner}.

\begin{figure*}
    \centering
    \plottwo{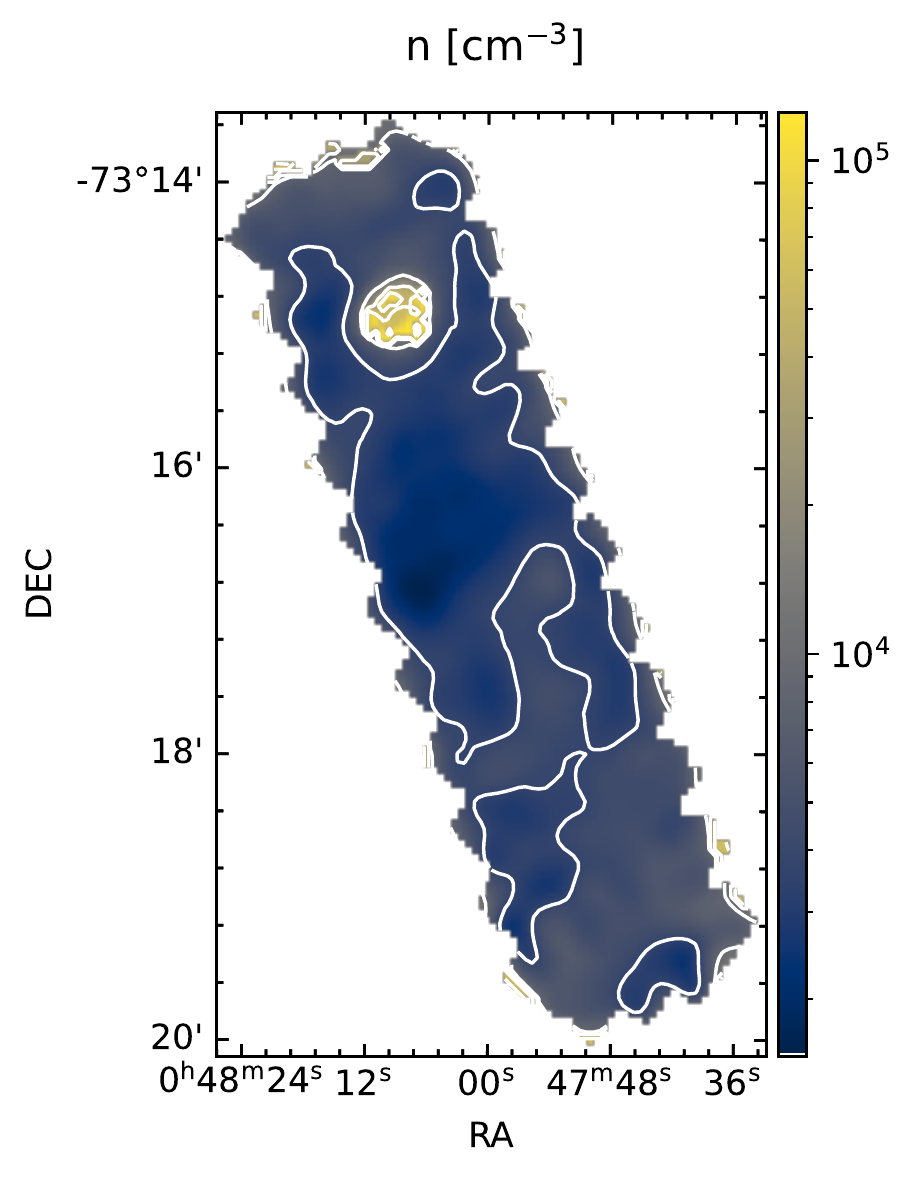}{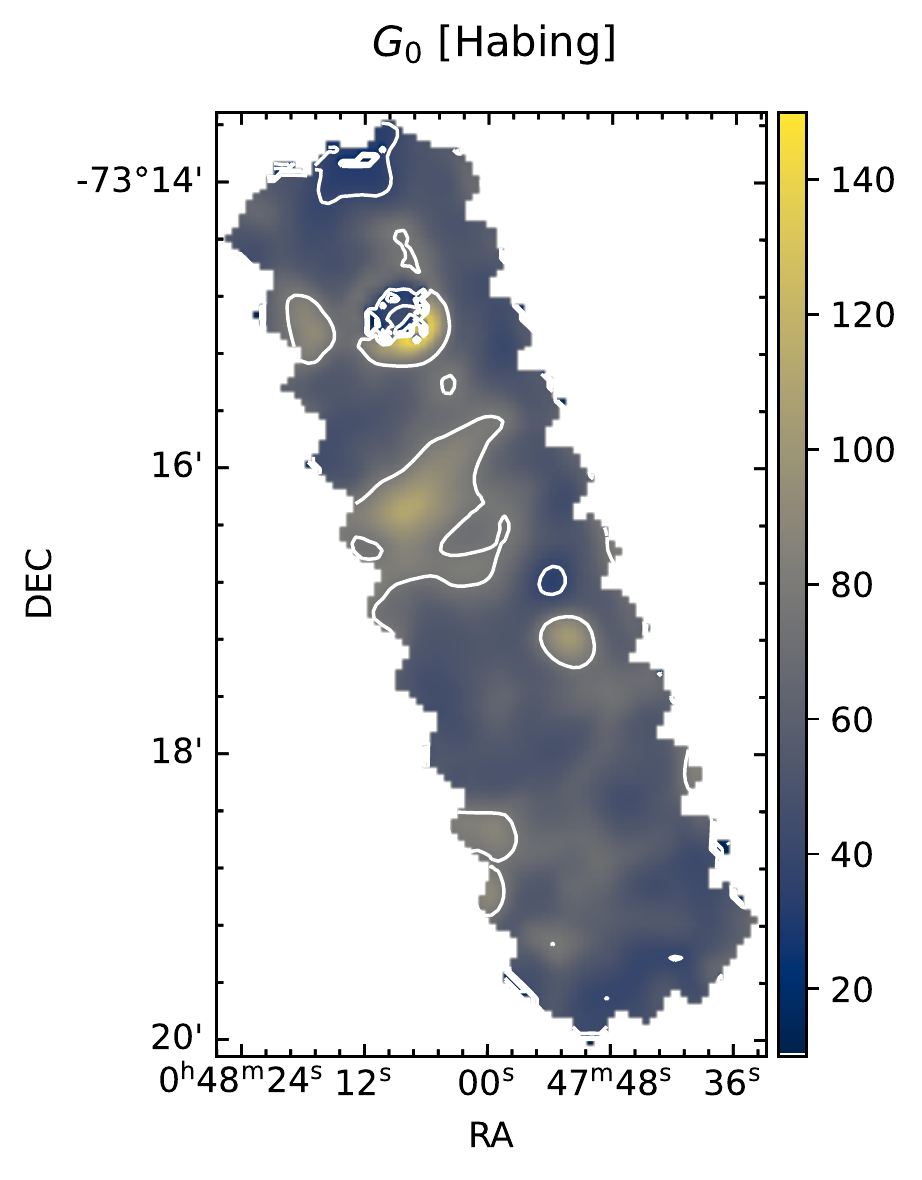}
    \caption{Maps of hydrogen nucleus density $n~{\rm (cm^{-3})}$ (left) and radiation field $G_0$ (Habing) (right) in the SMC star-forming cloud N22 as determined by {\tt LineRatioFit}. The fit uses a {\tt ModelSet} computed with the metallicity of the SMC (Z=0.3). See Listing \ref{listing4}.}
    \label{fig:mapfit}
\end{figure*}

One of the significant improvements that \pdrtpy\ makes over the old web version is the ability to operate on images, creating maps of best-fit $n$ and $F_{FUV}$.  Listing 4 \ref{listing4} (Figure \ref{fig:mapfit}) shows an example using the [C II]~158 \micron, [O I]~63 \micron, and far-infrared continuum maps in the Small Magellanic Cloud N22 star-forming region from \cite{Jameson2018}.  In this example, models computed using a low metallicity (Z=0.3) were used to match the conditions of the SMC.  To fit the 4768 non-blanked pixels takes $\sim 17$s in Jupyter notebook on a late-model laptop using a single CPU.  We have not yet implemented multi-threading speedups. The fit was done with the NNLS method;  using {\tt emcee} on so many pixels would be prohibitive (about 14s per pixel or over 18 hr for the entire map).

    \subsubsection{\protect $H_2$ Excitation Diagrams}
From the observed, extinction-corrected intensity $I$ of an 
${\rm H_2}$ spectral line we can calculate the the column density in the upper state, $N_u$:
\begin{equation}
N_u = 4\pi {I\over A\Delta E},
\end{equation}
where $A$ is the Einstein A coefficient and $\Delta E$ is the energy difference between the upper and lower states of the transition. More typically one is interested in the normalized upper state column density $N_u/g_u$ for each transition, where $g_u$ is the statistical weight of the upper state. 

An excitation diagram plots the upper state energy of the transition $E_u/k$ on the x-axis versus $log(N_u/g_u)$ on the y-axis.   The statistical weight $g_u = (2I+1)\times(2J+1)$, where $I$ is
the total nuclear spin and $J$ is the rotational quantum number of the upper level.
Ortho hydrogen molecules have the spins of both the nuclei in the same direction $I=1$, and odd $J$; para molecules have nuclei that spin in opposite directions $I=0$, and even $J$. In LTE at temperatures $T\gtrsim 200$~K, the ortho-to-para ratio (OPR) of molecules is in the ratio  of 3 to 1.   In non-equilibrium environments,  OPR can be less than 3, and the actual $N_u/g_u$ will increase over its LTE value (see discussions in \citealt{Burton1992} and \citealt{Sheffer2011}). In such cases,  on a traditional plot that assumes OPR=3, $N_u$ for odd $J$ will be measured as too low.  This creates the so-called ``zig-zag" pattern in the excitation diagram \citep{Neufeld1998, Fuente1999, Sternberg1999}.

Often, excitation diagrams show evidence of both ``hot" and ``cold" gas components, where the ``cold" gas dominates the intensity in the low $J$ transitions and the``hot" gas dominates in the high $J$ transitions, leading to a curved line in the diagram. 
Given data over several transitions, 
one can fit for $T_{cold}, T_{hot}$, $N_{total} = N_{cold}+ N_{hot}$, and OPR.  The $T_{cold}$ is usually a good approximation for
the gas kinetic temperature since the lower levels are collisionally excited. The $T_{hot}$ is a generally a result of UV fluorescence to the
excited levels. One needs at least 5 data points to fit the temperatures and column densities (slope and intercept $\times 2$), though one could compute (not fit) them with only 4 points. To additionally fit OPR requires 6 data points. The cold, hot, and total column densities, are computed using $N_0$ determined from y-axis intercepts and the partition function $Z(T) = 0.0247~T ~ [1 - {\rm exp}( -6000/T)]^{-1}$, where $T$ is one of $T_{cold}$ or $T_{hot}$ \citep{Herbst1996}.  As with $n$ and $F_{FUV}$ fitting,  the user can fit single pixels or full maps. For H$_2$ map inputs, PDRT will fit the excitation diagram at every pixel and produce maps of $T_{cold}, T_{hot}$, $N_{total}, N_{cold}, N_{hot}$, and OPR.  The method {\tt H2ExcitationFit.explore} lets users interactively probe the resultant maps and excitation fits.

\begin{figure}
    \centering
    \includegraphics[width=\columnwidth]{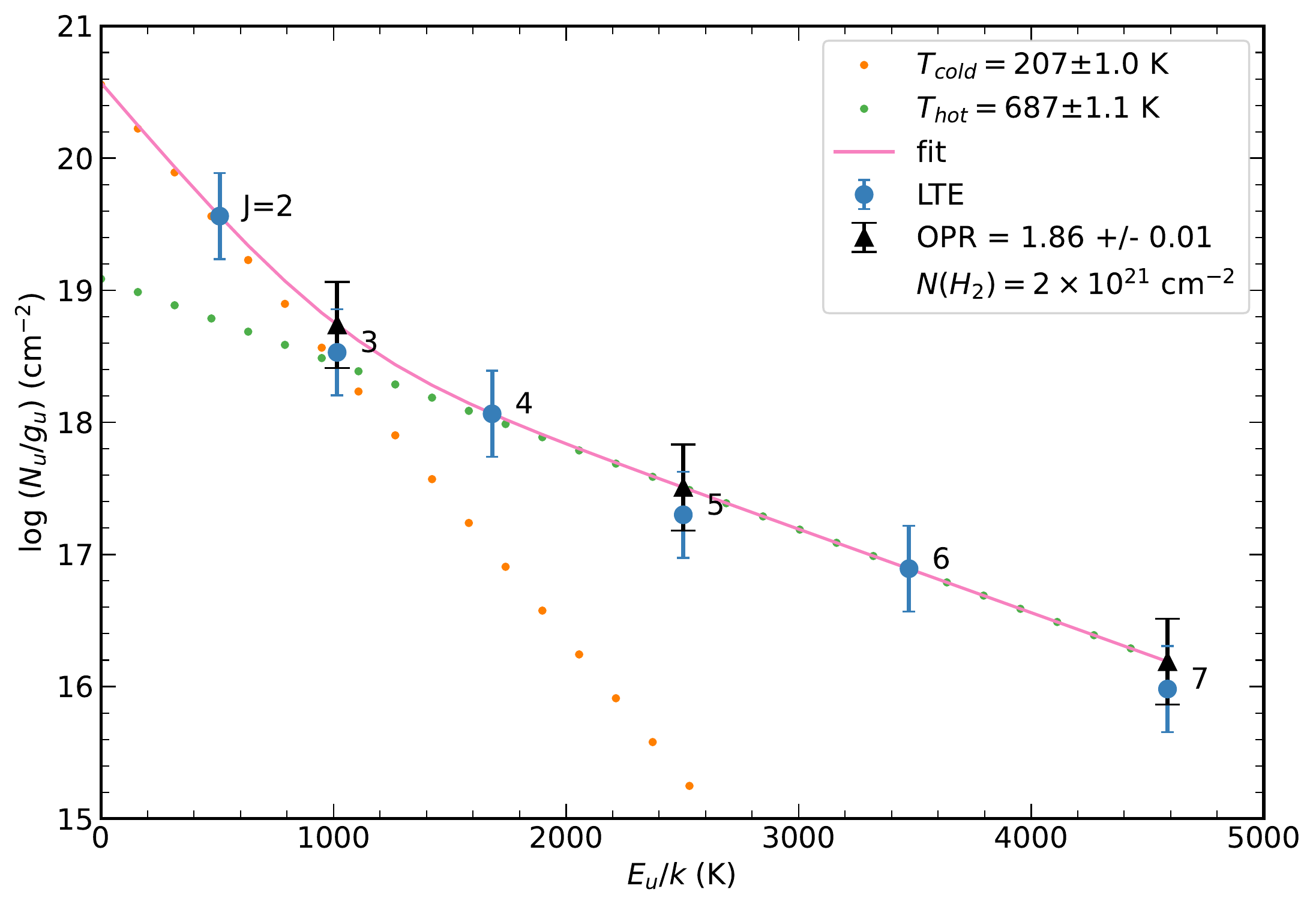}
   \caption{Example of PDRT fitting of a molecular hydrogen rovibrational line excitation diagram to determine temperature and column density for both hot and cold gas components, total column density $N({\rm H_2})$, and ortho-to-para ratio (OPR). $E_u/k$ is the upper state energy of the transition and $N_u/g_u$ is the normalized upper state column density. Blue dots are data from \cite{Sheffer2011} and black triangles are the data adjusted for the fitted OPR.  See Listing \ref{listing5}}
    \label{fig:h2opr}
\end{figure}

Listing \ref{listing5} gives an example of fitting the excitation conditions given observations of six ${\rm H_2}$ rotational lines.  The data are of NGC 2023 and taken from Figure 9 of \cite{Sheffer2011}, adding an artificial point for the J=6 line to allow for fitting of OPR.   Figure \ref{fig:h2opr} shows the result of the fit.  We find the same temperatures and OPR within the errors as \cite{Sheffer2011}.

For both {\tt LineRatioFit} and {\tt H2ExcitationFit},  the fit results are stored per pixel in an {\tt FitMap} object which derives from {\tt astropy.nddata.NDData}. The {\tt FitMap will } contains {\tt lmfit.model.ModelResult}
objects for {\tt H2ExcitationFit} or {\tt lmfit.minimizer.MinimizerResult} objects {\tt LineRatioFit}.  The user can thus examine in detail the fit at any pixel.

\subsection{H II Region Diagnostics}

Observations of fine-structure line ratios arising in 
ionized gas can be used to estimate the electron density, $n_e$, and gas temperature, $T_e$, in an H II region. In general, lines that arise from different energies above ground give
estimates of the gas temperature, while lines that arise from similar levels above ground but with different collision strengths give estimates of the gas density. 
We focus on lines that are expected to be bright in JWST observations of Galactic H II regions, namely those arising from ${\rm Fe^+}$, 
${\rm Ar^{+2}}$, and ${\rm Ar^{+4}}$. In particular, 
${\rm Fe^+}$ has great potential for producing diagnostic line ratios due to the large number of levels excited in 
an H II region, but with  caveats as noted here. 
Low level
${\rm Fe^+}$ line emission is also produced in the
neutral gas within the PDR and thus the same species
could trace physical conditions continuously from ionized
to neutral gas. 
   
\begin{figure}
    \centering
    \includegraphics[width=\columnwidth]{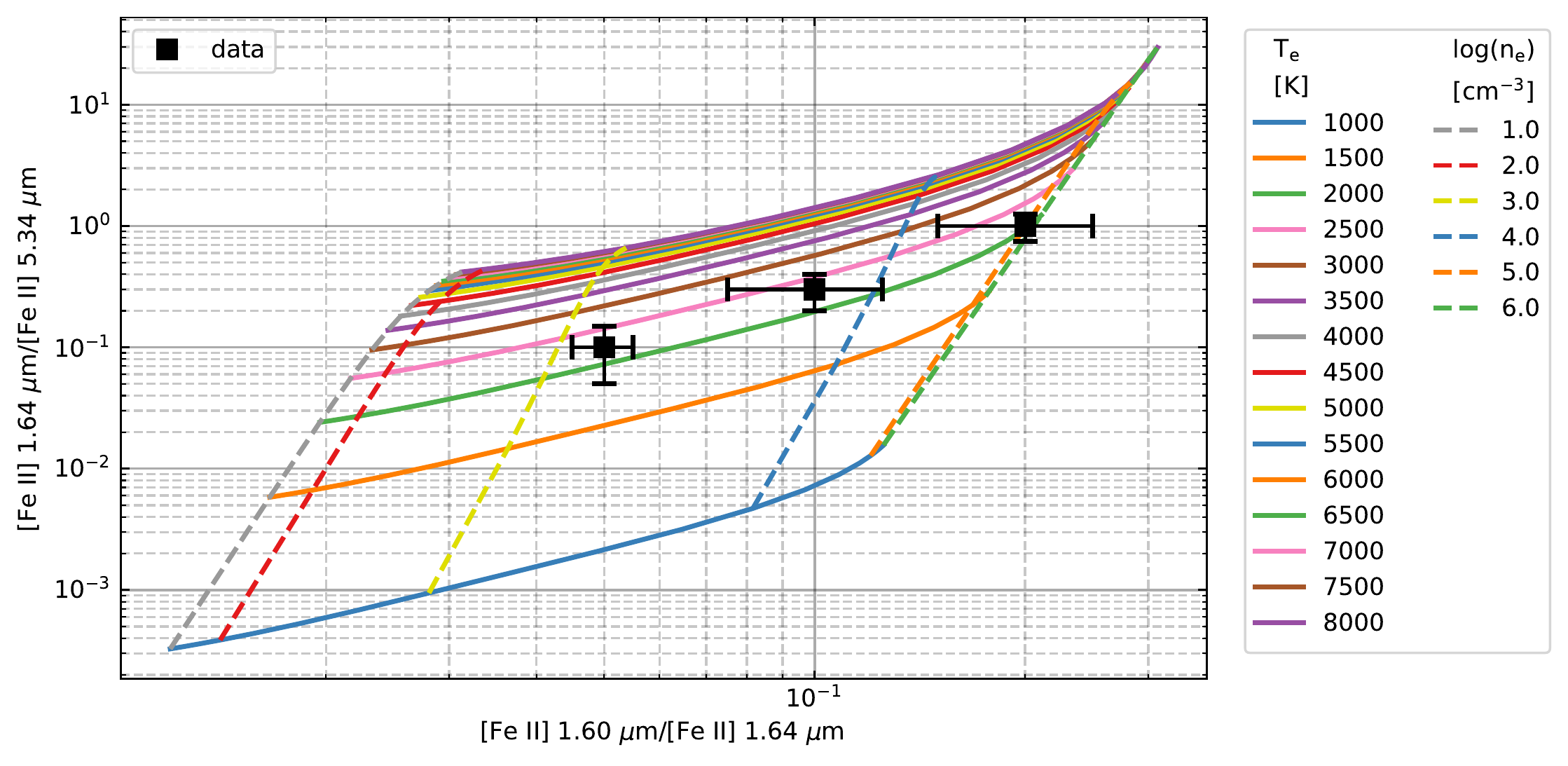}
   \caption{Example of PDRT diagnostic plot for electron density, $n_e$, and gas temperature, $T_e$, from the [Fe II] line ratios [Fe II] 1.64 $\mu$m/[Fe II] 5.34 $\mu$m versus
   [Fe II] 1.60 $\mu$m/[Fe II] 1.64 $\mu$m.
  Squares are sample data points with error bars.  See Listing \ref{listing1}.
   }
    \label{fig:HIIdiagnostic}
\end{figure}

We assume that the line emission is in the optically thin
limit so that the ratio of intensities is given by 
the ratio of volume emissivity. For ${\rm Ar^{+2}}$ and 
${\rm Ar^{+4}}$ we use CHIANTI \citep{Dere1997, DelZanna2021} using the default values
for the A values and collision strengths.
For ${\rm Fe^+}$, we substituted the default values in 
CHIANTI with A values from \cite{Deb2011} and collision strengths from \cite{Smyth2019}. 
The emissivity ratios are found in the temperature range
from $T_e =10^3$ K to $10^4$ K, and the density range from 
$n_e = 10^2$ ${\rm cm^{-3}}$ to $10^6$ ${\rm cm^{-3}}$.
Fits files are constructed of the resulting values,
and phase space plots and data overlays can then be made using 
the tools discussed in subsection \ref{subsec:Data}.
A sample figure is shown in Figure \ref{fig:HIIdiagnostic}.
We note, however, that for [Fe II] fine-structure lines the published A values and collision strengths vary between different authors and in some cases do not agree with
the observations
\cite[e.g.,][]{Koo2016} so the
[Fe II] plots must be considered tentative until the atomic
data can be further verified by observations, laboratory work, 
or quantum calculations.

    \section{Future development}\label{sec:future}
Below we describe a few future enhancements to the PDR model code, model data, and analysis tools that we intend to undertake.  This is not an exhaustive list; we encourage users to submit other requests via github.  We also encourage developers to pitch in!
\subsection{Changes to Wolfire-Kaufman model code}
\subsubsection{Additional viewing angles}
The Wolfire-Kaufman PDR models are designed to predict emission line intensities for a
face-on geometry in which the line-of-sight and
direction of illumination are parallel.
However, there are many PDRs that are
observed edge-on in which the line-of-sight and direction of illumination are perpendicular and the
layers of the PDR are spread across the sky (e.g., Orion Bar). The line intensities are then a function of position. The peak line intensities and dust continuum
can either increase or decrease significantly, compared to face-on, depending on the layer thickness along the line of sight.
We will add grids of edge-on models following the prescription given in 
\citep{Pabst2017} which accounts for optical depth effects in the lines. Using a similar technique we will also provide emitted line intensities for 
a viewing angle of 45 degrees.  Having the three cases, face-on, edge-on, and 45 degrees, will help users understand better the geometries of their sources and how viewing angle can affect the observed intensities.

\subsubsection{Deuterium chemistry}
The lowest HD rotational line at 112 $\mu$m is much easier to excite than
${\rm H_2}$ due to its 4 times lower energy above ground. 
When combined with PDR models,  HD  provides a particularly good measure 
of the warm molecular gas mass as well as the D/H ratio which is important for cosmological simulations. It is also a  prime target in protostellar disk 
observations. We will add deuterium chemistry to the PDR model using a simple network \citep{LePetit2002} with reaction rates updated, for example, from
KIDA\footnote{\url{https://kida.astrochem-tools.org}} and collision rates updated from \cite{Wan2019}. The output will be the IR  rotational and vibrational line emission as a function of $n$ and $F_{FUV}$.

\subsubsection{Improved treatment of metallicities}
A large database of PDR observations are available covering a range of metallicities. For example, the Herschel KINGFISH survey \citep{Kennicutt2011} mapped galaxies with 
a metallicity range of 0.04 to 5 relative to solar while SOFIA and Herschel have
mapped regions in the LMC (0.5 solar) \citep{Lebouteiller2019, Okada2019, Chevance2016, Lee2016, Lee2019} and 
the SMC (0.2 solar) \citep{RequenaTorres2012, Jameson2018}.
Although traditionally modelers have used the same 
scaling for dust abundance (responsible for extinction and heating), and metals
(responsible for gas cooling) 
it is now known that these scalings diverge for metallicity $Z < 0.2$ relative to the local ISM \citep{RemyRuyer2014}.  
We will provide a series of models that covers the metallicity range 0.03-5, while accounting for the expected scaling in dust and metals.
\subsection{Changes to \pdrtpy\ code}
\subsubsection{Correction for [O I] and [C II] absorption.}
It has become increasingly apparent that the [O I] 63 $\mu$m, and [C II] 158 $\mu$m lines can be self-absorbed, thereby affecting the
interpretation of the integrated line intensities \citep{Graf2012,Guevara2020,Goldsmith2021,Kabanovic2022,Bonne2022}. This is most often noticed by a
[OI] 145 $\mu$m/63 $\mu$m ratio greater than 0.1 or
a central dip in velocity resolved profiles seen in one isotope but not another (e.g., [$^{12}$C II] vs.\ [$^{13}$C II]). Note that PDR models use an escape probability formalism for the line transfer that accounts for the absorption within the PDR but not for cold foreground absorption. It is not known {\em a priori} what the correction for absorption  should be and observers usually adopt a correction factor of 2-3x increase in the observed [O I] line intensity so that other line ratios are physically realistic \citep[e.g.,][]{Schneider2018,Goldsmith2021}. We will assist 
users by providing plots of appropriate correction factors to use for both the [O I] 63 $\mu$m and 
[C II] 158 $\mu$m lines as functions of the foreground gas temperature and column density and as functions of 
the line center optical depth. We realize these are not
perfect solutions but they do provide a better understanding of the source environment and direct users to physically motivated solutions.

\subsubsection{Regularization for image-based fitting} 
\begin{figure}[ht]
    \centering
        \includegraphics[width=0.85\columnwidth]{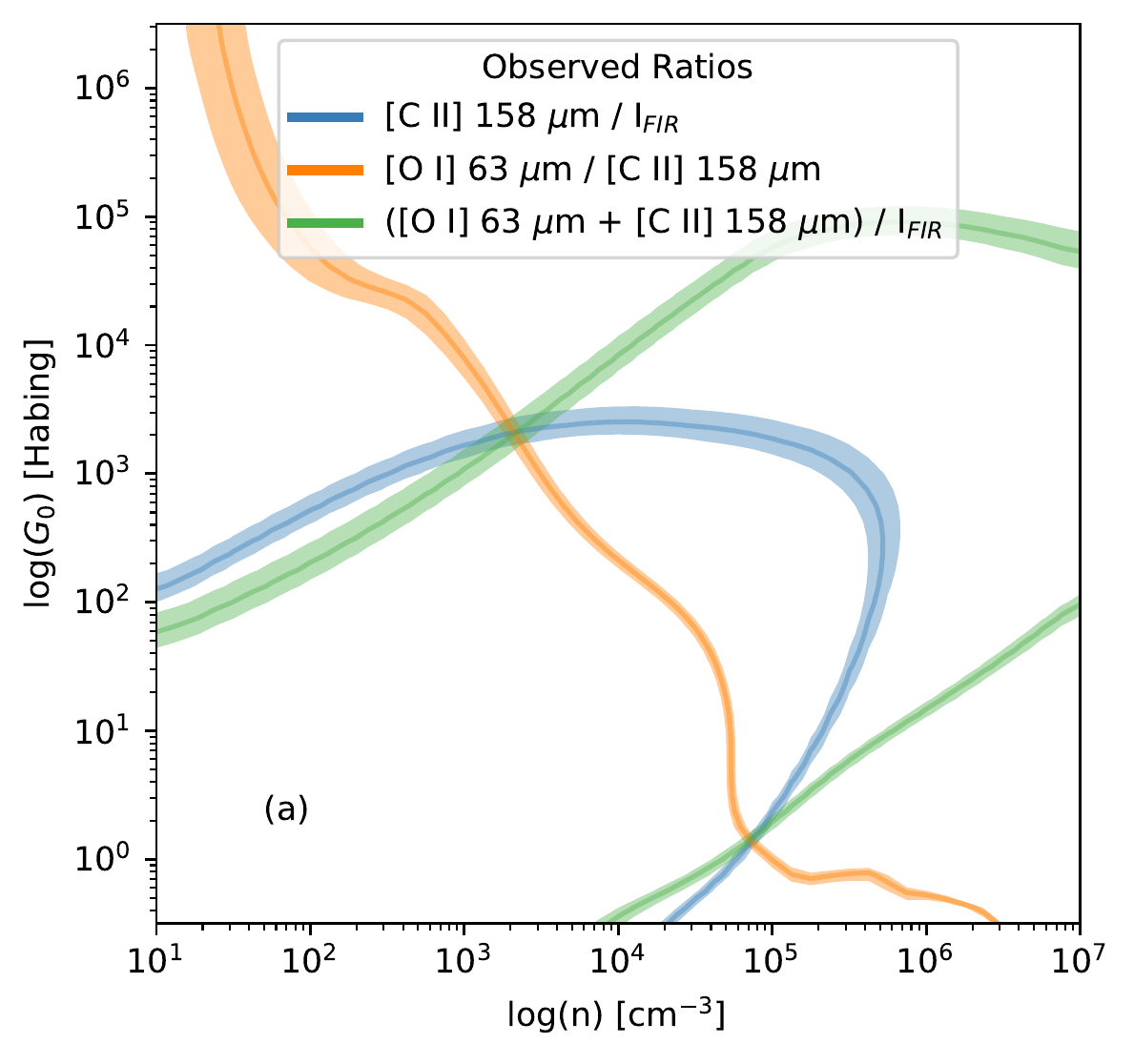}\\
        \includegraphics[width=\columnwidth]{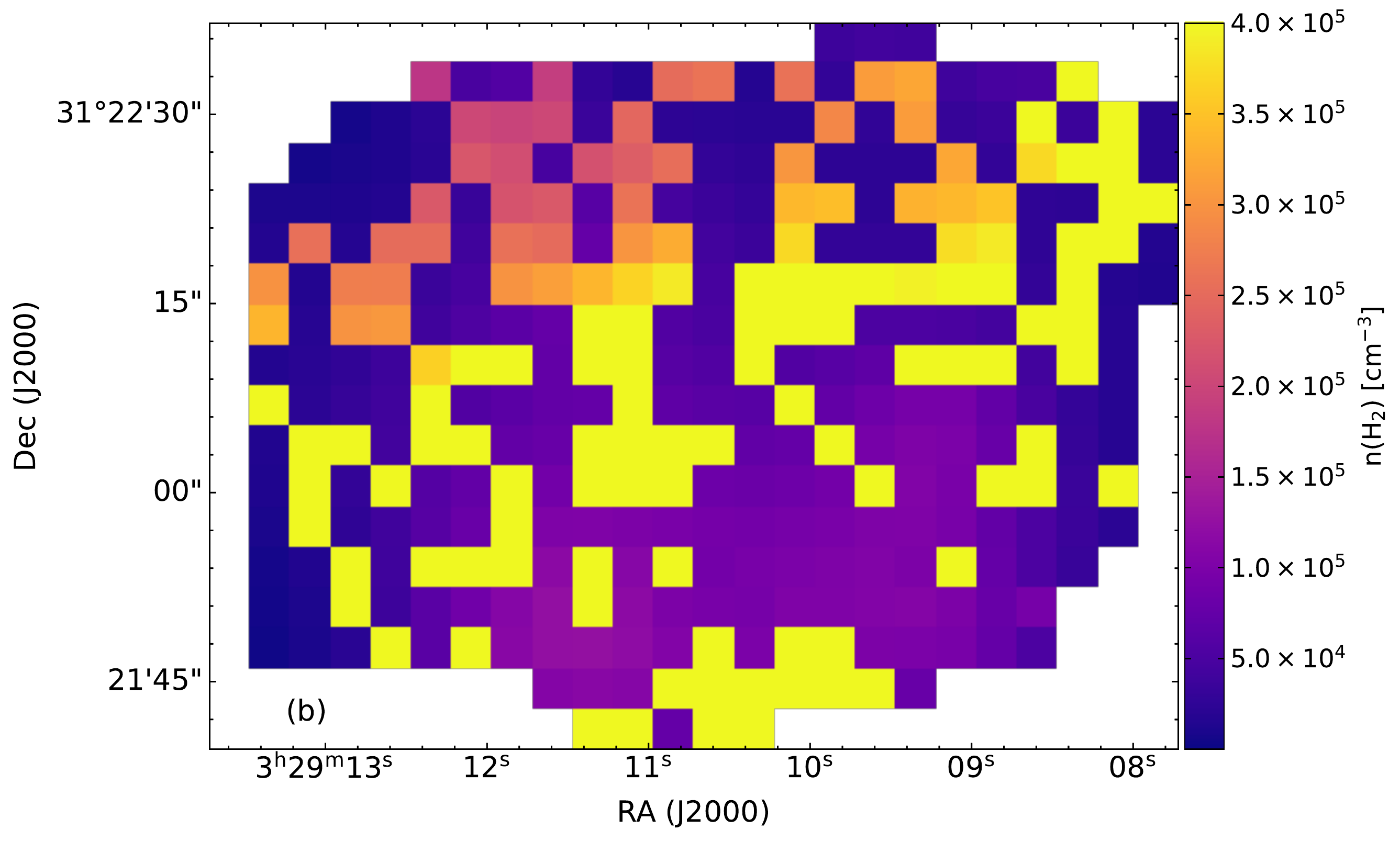}\\
        \includegraphics[width=\columnwidth]{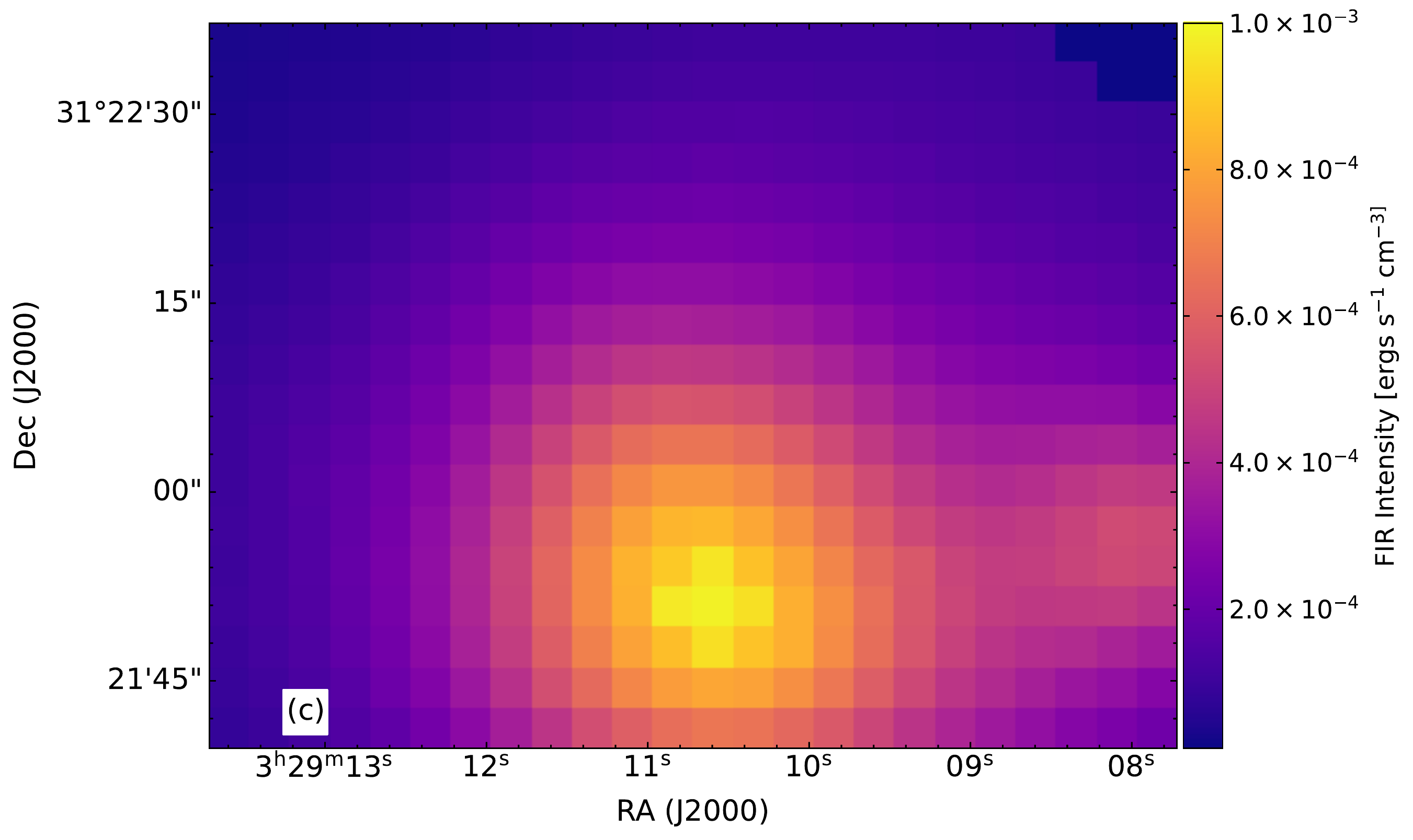}
    \caption{Example of when fitted solutions get into trouble.  {\it (a)} Overlay diagram in model space ($n,G_0$) of ratios of 3 common observations: [C II], [O I], and FIR intensity; the width of lines indicate observational uncertainties.  Potentially valid solutions appear at crossing points in two very different locations of model space. {\it (b)} Least-squares fitted density map of NGC 1333 from observations. Mirroring the behavior of {\it (a)}, adjacent pixels can have very different derived densities, despite that the observations {\it (c)} are smooth on a larger spatial scale.}
    \label{fig-twosolutions}
\end{figure}

When determining best-fit density and radiation field maps in instances where the observations are few or are of low S/N, it is possible for a fitting algorithm to oscillate between two nearly equally good solutions (Figure \ref{fig-twosolutions}).  The best way to break this degeneracy would be to obtain additional observations that further constrain the solution, but this is often not practicable, so another method is needed.

When a fitted solution depends discontinuously upon the initial data, as in our example, it is a symptom of an ill-posed (or ill-conditioned) problem.  A common method to resolve an ill-posed problem is to employ a regularization technique \citep{Tikhonov1977}.  Simply put, regularization means replacing the problem with a different problem whose solution roughly matches the desired solution (has low bias), is less sensitive to noise (has low variance), and has a parameter that allows bias-variance tradeoff.  

One way to do this is to add a ``penalty'' to the usual minimization function  $M$  that can be used as an additional constraint:
\begin{equation}
    M = \underbrace{%
    \sum_{i}^{N}{{[ y_i^{\rm obs} - y_i^{\rm model}({\rm \textbf{v}}) ]^2} \over
    {\sigma_i^2}}}_\mathrm{\chi^2} 
    + \underbrace{\vphantom{\sum_{i}^{N}}\lambda [f_i({\rm\textbf{v}})]^2}_\mathrm{ penalty}
\end{equation}
\noindent where $y_i^{\rm obs}$
 is the set of observed data, 
 $y_i^{\rm model}$ is the model calculation,  \textbf{v} is the set of variables in the model to be optimized in the fit,  $\sigma_i$
 is the estimated uncertainty in the data, and $\lambda$ is the bias-variance tuning parameter.
 
Because observed maps are typically smooth over several pixels (e.g., Figure \ref{fig-twosolutions}{\it (c)}), a reasonable choice would be a penalty function that enforces local smoothness.  For instance, an inverse-distance weighted sum of the spatial derivatives of the observed maps could act as a penalty (``the solution cannot be less smooth locally than the observations").  The choice of $\lambda$ is key and can vary with different sets of observations.  
Another regularization technique is to characterize the spatial structure of the observational and fitted solution maps with a wavelet transform \citep{Mallat2012, Allys2019} and to favor solutions that have the wavelet components found in the observations.  This method can be computationally quick and has been successful in medical imaging \citep{Guerquin2009wavelet}.

We will explore different regularization methods, test them against real and simulated data, and implement those that perform well (e.g.,  low-to-modest computational cost, few corner cases, robust to adding or subtracting data). We will provide guidance to users on choosing $\lambda$ as well as explore ways to have PDRT choose it for them, for instance, by examining the Akaike Information Criterion \citep{Burnham2003} which is already calculated for every pixel by PDRT.  The regularization enhancement will also apply to the ${\rm H_2}$ excitation fitting tool in its map mode.   

\section{Summary}

The PDR Toolbox is a mature, versatile package for analysis of photodissociation regions with a wide range of physical conditions. It is applicable to observations from many telescopes from the IR to the sub-mm regimes.  It can also be used to compare models from different PDR codes. 

\begin{acknowledgements}
We are grateful for support from the NASA Astrophysics Data Analysis Program award \#80NSSC19K0573, from the 
SOFIA Legacy Program, FEEDBACK, provided by NASA
through award SOF070077 issued by USRA to the University of Maryland, and from JWST-ERS program ID 1288 
through a grant from the Space Telescope Science Institute under
NASA contract NAS5-03127 to the University of Maryland.
We are grateful for helpful conversations with and manuscript comments from the FEEDBACK and PDRs4All teams. Special thanks to Maitraiyee Tiwari for significant beta-testing.  We thank Markus R{\"o}llig and Volker Ossenkopf-Okada for providing the KOSMA-tau models.
Support for the early development of the PDRT CGI version came from NASA Astrophysics Data Program and Applied Information Systems Research Program grants.  We thank the anonymous referee for helpful comments.
\end{acknowledgements}
\clearpage
\appendix
\section{Code Listings}
\subsection{Models, Modelsets, and ModelPlot}\label{listing1}
\begin{lstlisting}[language=Python]
###########################################################################
###            Listing A.1:  models, ModelSet, and ModelPlot            ###
###########################################################################

from pdrtpy.modelset import ModelSet
from pdrtpy.plot.modelplot import ModelPlot
from pdrtpy.measurement import Measurement
import pdrtpy.pdrutils as utils
import astropy.units as  u
from astropy.nddata import StdDevUncertainty

# Get the Wolfire-Kaufman 2020 Z=1 models 
ms = ModelSet("wk2020",z=1)

# Get KOSMA-tau R=3.1 model
mskt = ModelSet("kt2013wd01-7",z=1,mass=100,medium='clumpy')

# Example of how to fetch a given model, the [OI] 63 micron/[CII] 158 micron intensity ratio.
# The returned model type is pdrtpy.measurement.Measurement.

model = ms.get_model("OI_63/CII_158")
modelkt = mskt.get_model("OI_63/CII_158")

# Find all the models that use some combination of CO(J=1-0), [C II] 158 micron, 
# [O I] 145 micron,  and far-infrared intensity. This example gets both intensity 
# and ratio models, though one can specify model_type='intensity' 
# or model_type='ratio' to get one or the other. 
# The models are returned as a dictionary with the keys set to the model IDs.

mods = ms.get_models(["CII_158","OI_145", "CO_10", "FIR"],model_type='both')
print(list(mods.keys()))
# Output of above: ['OI_145', 'CII_158', 'CO_10', 'CII_158/OI_145', 'CII_158/CO_10', 
#                   'CII_158/FIR', 'OI_145+CII_158/FIR']

# Plot a selected model and save it to a PDF file.  Note in this example,
# we request Habing units for the FUV field.

# WK
mp = ModelPlot(ms)
mp.plot('OI_145+CII_158/FIR',yaxis_unit='Habing',
        label=True, cmap='viridis', colors='k',norm='log')
mp.savefig("example1a_figure.pdf")
# KT
mpkt = ModelPlot(mskt)
mpkt.plot('OI_145+CII_158/FIR',yaxis_unit='Habing',
        label=True, cmap='viridis', colors='k',norm='log')
mpkt.savefig("example1b_figure.pdf")

rcw49 = []
label = ["shell","pillar","northern cloud","ridge"]
format_ = ["k+","b+","g+","r+"]
# The data files are in the testdata directory of the pdrtpy installation
for region in ["shell","pil","nc","ridge"]:
    f1 = utils.get_testdata(f"cii-fir-{region}.tab")
    f2 = utils.get_testdata(f"cii-co-{region}.tab")
    rcw49.append(Measurement.from_table(f1))
    rcw49.append(Measurement.from_table(f2))
    
mp.phasespace(['CII_158/FIR','CII_158/CO_32'],nax1_clip=[1E2,1E5]*u.Unit("cm-3"),
               nax2_clip=[1E1,1E6]*utils.habing_unit, measurements=rcw49,label=label,
               fmt=format_,title="RCW 49 Regions")
mp.savefig("example1c_figure.pdf")

# Example ionized gas line diagnostic diagram
i1 = Measurement(identifier='FEII_1.60/FEII_1.64',
                 data=[0.1,0.05,0.2],
                 uncertainty=StdDevUncertainty([0.025,0.005,0.05]),unit="")
i2 = Measurement(identifier='FEII_1.64/FEII_5.34',
                 data=[0.3,0.1,1.0],
                 uncertainty=StdDevUncertainty([0.1,0.05,0.25]),unit="")
mp.phasespace(['FEII_1.60/FEII_1.64','FEII_1.64/FEII_5.34'],
              nax2_clip=[10,1E6]*u.Unit("cm-3"),nax1_clip=[1E3,8E3]*u.Unit("K"),
             measurements=[i1,i2],errorbar=True)
mp.savefig("example1d_figure.pdf")
\end{lstlisting}
\clearpage

\subsection{ Fitting intensity ratios for single-pixel observations}\label{listing2}
\begin{lstlisting}[language=Python]
#############################################################################
###  Listing A.2: Fitting intensity ratios for single-pixel observations  ###
#############################################################################
from pdrtpy.measurement import Measurement
from pdrtpy.modelset import ModelSet
from pdrtpy.tool.lineratiofit import LineRatioFit
from pdrtpy.plot.lineratioplot import LineRatioPlot
import pdrtpy.pdrutils as utils
from astropy.nddata import StdDevUncertainty
from lmfit import fit_report

# Example using single-beam observations of [OI] 163 micron, [CI] 609 micron, CO(J=4-3), 
# and [CII] 158 micron lines. You create `Measurements` for these using the constructor
# which takes the value, error, line identifier string, and units.   The value and the error 
# must be in the same units.  You can mix units  in different Measurements; note we use 
# K km/s  for the CO observation below.   The PDR Toolbox will convert all `Measurements` 
# to a common unit before using them in a fit. You can also add optional beam size 
# (bmaj,bmin,bpa), however the tools requires all `Measurements` have the same beam size 
# before calculations can be performed.  (If you don't provide beam parameters for any of 
# your Measurements,  the Toolbox will assume they are all the same).

myunit = "erg s-1 cm-2 sr-1" # default unit for value and error
m1 = Measurement(data=3.6E-4,uncertainty = StdDevUncertainty(1.2E-4),
                 identifier="OI_63",unit=myunit)
m2 = Measurement(data=1E-6,uncertainty = StdDevUncertainty([3E-7]),
                 identifier="CI_609",unit=myunit)
m3 = Measurement(data=26,uncertainty = StdDevUncertainty([5]),
                 identifier="CO_43",restfreq="461.04077 GHz", unit="K km/s")
m4 = Measurement(data=8E-5,uncertainty = StdDevUncertainty([8E-6]),
                 identifier="CII_158",unit=myunit)
observations = [m1,m2,m3,m4]

ms = ModelSet("wk2020",z=1)

# Instantiate the LineRatioFit tool giving it the ModelSet and Measurements
p = LineRatioFit(ms,measurements=observations) 
p.run()
# Print the fitted quantities using Python f-strings and the fit report via lmfit
print(f"n={p.density:.2e}\nX={utils.to('Draine',p.radiation_field):.2e}")
print(fit_report(p.fit_result[0]))

# Create the plotting tool for the LineRatioPlot, 
# then make plots of the observed ratios overlayed on the model ratios
plot = LineRatioPlot(p)
plot.ratios_on_models(yaxis_unit="Draine",colorbar=True,norm='log',
                      cmap='cividis',label=True,ncols=3,figsize=(23,7))
plot.savefig('example2_figure.pdf')

plot.overlay_all_ratios(yaxis_unit="Draine",figsize=(6,7))
plot.savefig("example3_figure.pdf")
# Plot the reduced chisquare, with only contours and legend
plot.chisq(image=False,colors='k',label=True,legend=True,yaxis_unit='Draine',figsize=(6,7))
plot.savefig("example4_figure.pdf")
\end{lstlisting}
    \clearpage
\subsection{Fitting intensity ratios for single-pixel observations with MCMC }\label{listing3}
\begin{lstlisting}[language=Python]
#######################################################################################
###  Listing A.3:  Fitting intensity ratios for single-pixel observations with MCMC ###
#######################################################################################

from pdrtpy.measurement import Measurement
from pdrtpy.modelset import ModelSet
from pdrtpy.tool.lineratiofit import LineRatioFit
from pdrtpy.plot.lineratioplot import LineRatioPlot
import pdrtpy.pdrutils as utils
from astropy.nddata import StdDevUncertainty
from copy import deepcopy
import corner
import numpy as np


myunit = "erg s-1 cm-2 sr-1" # default unit for value and error
m1 = Measurement(data=3.6E-4,uncertainty = StdDevUncertainty(1.2E-4),
                 identifier="OI_63",unit=myunit)
m2 = Measurement(data=1E-6,uncertainty = StdDevUncertainty([3E-7]),
                 identifier="CI_609",unit=myunit)
m3 = Measurement(data=26,uncertainty = StdDevUncertainty([5]),
                 identifier="CO_43",restfreq="461.04077 GHz", unit="K km/s")
m4 = Measurement(data=8E-5,uncertainty = StdDevUncertainty([8E-6]),
                 identifier="CII_158",unit=myunit)
observations = [m1,m2,m3,m4]

ms = ModelSet("wk2020",z=1)

# Instantiate the LineRatioFit tool giving it the ModelSet and Measurements
p = LineRatioFit(ms,measurements=observations) 
p.run(method='emcee',steps=2000)

res = p.fit_result[0]

# the value of the Draine unit in cgs
scale = utils.draine_unit.cgs.scale
# copy the results table
rescopy = deepcopy(res.flatchain)

# scale the radiation_field column of the table to Draine since it is in cgs
rescopy['radiation_field'] /= scale # = np.log10(rescopy['radiation_field']/scale)
#rescopy['density'] = np.log10(rescopy['density'])
# now copy and scale the "best fit" values where the cross hairs are plotted.
truths=np.array(list(res.params.valuesdict().values()))
truths[1] /=scale
#truths = np.log10(truths)

fig = corner.corner(rescopy, bins=20,range=[(1E4,1.2E5),(10,500)],
                    labels=[r"$n~{\rm [cm^{-3}]}$",r"$\chi~{\rm [Draine]}$"],
                    truths=truths)
fig.savefig("example5_figure.pdf",facecolor='white',transparent=False)
\end{lstlisting}
     \clearpage
 \subsection{Fitting intensity ratios for map observations}\label{listing4}    
\begin{lstlisting}[language=Python]
############################################################################
###  Listing A.4:  Fitting intensity ratios for map observations         ###
############################################################################
from pdrtpy.measurement import Measurement
from pdrtpy.modelset import ModelSet
from pdrtpy.tool.lineratiofit import LineRatioFit
from pdrtpy.plot.lineratioplot import LineRatioPlot
import pdrtpy.pdrutils as utils

# Get the input filenames of the FITS files in the testdata directory
# utils.get_testdata() is a special method to locate files there.
# These are maps from Jameson et al 2018.
print("Test FITS files are in: %s"%utils.testdata_dir())
cii_flux = utils.get_testdata("n22_cii_flux.fits")  # [C II] flux
cii_err = utils.get_testdata("n22_cii_error.fits")  # [C II] error
oi_flux = utils.get_testdata("n22_oi_flux.fits")    # [O I] flux 
oi_err = utils.get_testdata("n22_oi_error.fits")    # [O I] error
FIR_flux = utils.get_testdata("n22_FIR.fits")       # FIR flux

# Output file names
cii_combined = "n22_cii_flux_error.fits"
oi_combined = "n22_oi_flux_error.fits"
FIR_combined = "n22_FIR_flux_error.fits"

# create the Measurements and write them out as FITS files with two HDUs.
Measurement.make_measurement(cii_flux, cii_err, 
                             outfile=cii_combined, overwrite=True)
Measurement.make_measurement(oi_flux, oi_err, 
                             outfile=oi_combined, overwrite=True)
# Assign a 10% error in FIR flux
Measurement.make_measurement(FIR_flux, error='10%',
                             outfile=FIR_combined, overwrite=True)

# Read back in the FITS files to Measurements
cii_meas = Measurement.read(cii_combined, identifier="CII_158")
FIR_meas = Measurement.read(FIR_combined, identifier="FIR")
oi_meas  = Measurement.read(oi_combined, identifier="OI_63")

# Here we will use the Small Magellanic Cloud ModelSet that have Z=0.1
# These are a limited set of models with just a few lines covered.
smc_ms = ModelSet("smc",z=0.1)
p = LineRatioFit(modelset=smc_ms, measurements=[cii_meas,FIR_meas,oi_meas])
p.run()
plot = LineRatioPlot(p)
plot.density(contours=True,norm="log",cmap='cividis')
plot.savefig("example6_n_figure.pdf")
plot.radiation_field(units="Habing",contours=True,norm="simple",cmap='cividis')
plot.savefig('example6_g0_figure.pdf')

# Save the results to FITS files.
p.density.write("N22_density_map.fits",overwrite=True)
p.radiation_field.write("N22_G0_map.fits",overwrite=True)
\end{lstlisting}
    \clearpage
        
   \subsection{Creating and fitting H2 excitation diagrams, including ortho-to-para ratio (OPR) }\label{listing5}      
\begin{lstlisting}[language=Python]
############################################################################
###  Listing A.5:  Creating and fitting H2 excitation diagrams,          ###
###                including ortho-to-para ratio (OPR)                   ###
############################################################################
from pdrtpy.measurement import Measurement
from pdrtpy.tool.h2excitation import H2ExcitationFit
from pdrtpy.plot.excitationplot import ExcitationPlot
from astropy.nddata import StdDevUncertainty
ntensity = dict()
intensity['H200S0'] = 3.003e-05
intensity['H200S1'] = 3.143e-04
intensity['H200S2'] = 3.706e-04
intensity['H200S3'] = 1.060e-03
intensity['H200S4'] = 5.282e-04
intensity['H200S5'] = 5.795e-04
observations = []
for i in intensity:
    m = Measurement(data=intensity[i],
                    uncertainty=StdDevUncertainty(0.75*intensity[i]),
                    identifier=i,unit="erg cm-2 s-1 sr-1")
    observations.append(m)

# Create the tool to run the fit
hopr = H2ExcitationFit(observations)
# Instantiate the plotter
hplot = ExcitationPlot(hopr,"H_2")

# Set some plot parameters appropriate for manuscript figure; 
# these pass through to matplotlib
hplot._plt.rcParams["xtick.major.size"] = 7
hplot._plt.rcParams["xtick.minor.size"] = 4
hplot._plt.rcParams["ytick.major.size"] = 7
hplot._plt.rcParams["ytick.minor.size"] = 4
hplot._plt.rcParams['font.size'] = 14
hplot._plt.rcParams['axes.linewidth'] =1.5
hplot.ex_diagram(ymax=21)
hplot.savefig('example9_figure.png',dpi=300)

# Fit a two temperature model allowing OPR to vary
hopr.run(fit_opr=True)
hplot.ex_diagram(show_fit=True,ymax=21)
hplot.savefig('example10_figure.png',dpi=300)
\end{lstlisting}

\bibliography{pdrt_ref.bib}
\software{Astropy \citep{astropy:2013, astropy:2018},
        emcee \citep{Foreman_Mackey_2013}, lmfit-py \citep{lmfit},
          matplotlib \citep{matplotlib2007CSE}, numpy \citep{numpy2011}
          }




\end{document}